\documentclass[a4paper,fleqn]{cas-dc}
\usepackage[utf8x]{inputenc} 
\usepackage[numbers]{natbib}
\usepackage{mathtools}
\usepackage{enumitem}

\usepackage{subcaption}

\usepackage[OT2, T1]{fontenc} 
\usepackage[russian, english]{babel}
\usepackage{graphicx}
\usepackage{todonotes}
\usepackage{amsfonts}
\usepackage{adjustbox}
\usepackage{booktabs}
\usepackage{makecell}
\usepackage{amsmath}
\usepackage{nicefrac}
\usepackage{comment}
\usepackage{pdflscape}
\usepackage{rotating}
\usepackage{multirow}
\usepackage[singlelinecheck=false]{caption}
\newtheorem{definition}{Definition} 

\newcommand{\approach}{\textit{LaSER}}

\newcommand{\lkg}{language-specific knowledge graph}
\newcommand{\cg}{language-specific click data}

\newcommand{\lkgCap}{Language-specific Knowledge Graph}
\newcommand{\cgCap}{Language-specific Click Data}
\newcommand{\ekgc}{Event\-KG+\-Click}

\newcommand{\voc}[2]{\texttt{#1:\allowbreak #2}}
\newcommand{\schema}[1]{\texttt{#1}}

\newcommand{\ed}[1]{\textcolor{black}{#1}}
\newcommand{\sa}[1]{\textcolor{black}{#1}}
\newcommand{\sg}[1]{\textcolor{black}{#1}}

\begin{document}


\title[mode = title]{LaSER: Language-Specific Event Recommendation}

\shorttitle{LaSER: Language-Specific Event Recommendation}    

\shortauthors{Sara Abdollahi, Simon Gottschalk and Elena Demidova}

\author[First]{Sara Abdollahi} [orcid=0000-0001-7752-146X]
\ead{abdollahi@L3S.de}

\author[First]{Simon Gottschalk}[orcid=0000-0003-2576-4640]
\ead{gottschalk@L3S.de}

\author[Second]{Elena Demidova}[orcid=0000-0002-5134-9072]
\ead{elena.demidova@cs.uni-bonn.de}

\address[First]{L3S Research Center, Leibniz Universität Hannover, Germany}

\address[Second]{Data Science \& Intelligent Systems Group (DSIS), University of Bonn, Germany}


\begin{abstract}
While societal events often impact people worldwide, a significant fraction of events has a local focus that primarily affects specific language communities. Examples include national elections, the development of the Coronavirus pandemic in different countries, and local film festivals such as the \textit{César Awards} in France and the \textit{Moscow International Film Festival} in Russia.
However, existing entity recommendation approaches do not sufficiently address the language context of recommendation. 
This article introduces the novel task of language-specific event recommendation, which aims to recommend events relevant to the user query in the language-specific context. This task can support essential information retrieval activities, including web navigation and exploratory search, considering the language context of user information needs.
We propose \approach{}, a novel approach toward language-specific event recommendation. 
\approach{} blends the language-specific latent representations (embeddings) of entities and events and spatio-temporal event features in a learning to rank model. This model is trained on publicly available Wikipedia Clickstream data. 
The results of our user study demonstrate that \approach{} outperforms state-of-the-art recommendation baselines by up to $33$ percentage points in MAP@5 concerning the language-specific relevance of recommended events.
\end{abstract}

\maketitle
\section{Introduction}
\label{sec:introduction}


Event and entity recommendation are critical tasks facilitating vital applications such as web navigation and exploratory research of a topic of user interest \cite{blanco2013entity}. Finding relevant events is an increasingly difficult task in the global digital world, where event relevance is highly dependent on the language context of the users and their information needs.  However, state-of-the-art event and entity recommendation approaches typically neglect this relevance dimension and provide results that do not adequately consider event-specific properties and language context.    

Table~\ref{tab:intro_example} provides examples of events particularly relevant to the Coronavirus pandemic from the perspective of the German, Italian, and Spanish-speaking audience on Wikipedia. This example is created based on the number of clicks in the Wikipedia Clickstream dataset~\cite{clickstream} that provides the click-through rates for the Wikipedia articles in the respective Wikipedia language edition. 
As we can observe, according to the clickstream, German Wikipedia users are mainly interested in the Coronavirus outbreaks in the German-speaking countries and the recession caused by the pandemic. Italian users are mainly interested in the pandemic in Italy and the SARS outbreak, a similar event in 2002, followed by the development in the US. The Spanish Wikipedia reflects user interests in several Spanish-speaking countries, such as Argentine, the US, and Colombia. These observations illustrate how event relevance varies based on the language-specific user context.

In this article, we present the new task of \textit{language-specific event recommendation}. 
This task adds two critical dimensions to entity recommendation: (i) recommendation of \textit{events} of societal importance, including political elections, military conflicts and sports events and (ii) the \textit{language-specific context} of these events. 
These dimensions are essential in various application scenarios, including event recommendation in information retrieval, event analytics to understand cultural viewpoints~\cite{gottschalk2018towards} and perception of events in different cultures~\cite{liu2005social}.

Language-specific event recommendations open up new web navigation and exploratory search opportunities and can assist users in researching events relevant to a topic in specific languages. Examples include historians researching the Second World War in different countries and journalists exploring the perception of the Coronavirus pandemic in different language communities. Language-specific event recommendations
can help address such information needs and provide recommendations that better fit users' interests and linguistic backgrounds.

\begin{table*}[ht]
\caption{Events with the highest number of clicks in the German, Italian and Spanish Wikipedia language editions starting from the article \textit{Coronavirus pandemic} in April 2021 (\textit{\#}: The number of user clicks in the respective Wikipedia language edition).}
\label{tab:intro_example}
\begin{tabular}{@{}rp{3cm}rp{3cm}rp{3cm}r@{}}
\toprule
\textbf{Rank} & \textbf{German} & \textbf{\#} & \textbf{Italian} & \textbf{\#} & \textbf{Spanish} & \textbf{\#} \\ \midrule
1 & COVID-19 pandemic in Germany & $3,775$ & COVID-19 pandemic in Italy & $2,890$ & COVID-19 pandemic in Argentina & $1,452$ \\ \addlinespace[0.2cm]
2 & COVID-19 recession & $1,852$ & 2002–2004 SARS outbreak & $780$ & COVID-19 pandemic in the United States & $1,286$ \\ \addlinespace[0.2cm]
3 & COVID-19 pandemic in Austria & $1,072$ & COVID-19 pandemic in the United States & $489$ & COVID-19 pandemic in Colombia & $1,105$ \\ \bottomrule
\end{tabular}
\end{table*}

The most relevant task addressed in the literature in the context of this article is entity recommendation, typically defined as the problem of suggesting entities relevant in a particular context, mostly provided as an entity of interest \cite{ni2020layered}. Entity recommendation has been tackled from different perspectives, including time-aware entity recommendation \cite{zhang2016probabilistic} and personalized recommendation of social events~\cite{khrouf2013hybrid}. Focusing on the language-specific event recommendation, we add novel dimensions to this task. We go beyond existing approaches that do not consider language-specific aspects and mainly optimize for entity popularity in general. Furthermore, in contrast to existing work, we train our model on publicly available data instead of relying on proprietary click or search logs typically used in the literature (e.g., \cite{ni2020layered}, \cite{huang2018learning}), enhancing the reproducibility of our results. 

In this article, we present \approach{} -- a new method for \textbf{La}nguage-\textbf{S}pecific \textbf{E}vent \textbf{R}ecommendation. Given a query entity of user interest (e.g., a person like \textit{Winston Churchill}, an event like  \textit{Coronavirus pandemic}, or a concept like \textit{Film Festival}) and a language of interest, \approach{} returns a list of events relevant to the query entity and the language community. With \approach{}, we tackle two key challenges of the language-specific event recommendation:

\begin{enumerate}[label=(C\arabic*)]
\itemsep0em

\item 
The creation of methods for language-specific event recommendation requires consideration of the language context,  
including latent properties of events and their relations in this context, along with the spatial and temporal dimensions. 
To the best of our knowledge, language-specific context has not been considered in state-of-the-art entity recommendations.

\item 
Training and evaluation of the models for language-specific event recommendation require corpora reflecting events users consider relevant in the specific language context. However, existing corpora do not contain information regarding language-specific user needs. Furthermore, datasets used for training recommendation models are often proprietary (e.g., the Yahoo! search logs~\cite{ni2020layered} and Baidu Web search engine logs~\cite{huang2018learning}). Recommendation methods based on such proprietary corpora are barely reproducible.
    
\end{enumerate}

To tackle these challenges in \approach{}, we derive and utilize event-specific and language-specific characteristics and include them in a language-specific recommendation model (C1), use freely available datasets, and collect high-quality user relevance judgments (C2).
\approach{} is based on language-specific latent representations (embeddings) of entities and events in a \lkg{} representing the relevance of entity and event relations in different language contexts. 
We combine these latent representations with spatio-temporal event features and utilize them for training a learning to rank (LTR) model. Given a language of interest and a query entity, this model generates a ranked list of relevant events. We train the model using the publicly available Wikipedia Clickstream.

We evaluate the effectiveness of \approach{} in two different setups. First, we evaluate the \approach{} ability to predict language-specific clicks between entities and events in the Wikipedia Clickstream. 
The results demonstrate that our model outperforms link-based, embedding-based and graph attention network based ranking baselines by over $8$ (nDCG@10) and $17$ (MAP@10) percentage points on average. 
Second, we conduct a user study to evaluate the relevance of recommended events and analyze different relevance criteria. The results confirm that \approach{} outperforms the baselines by up to $33$ percentage points in MAP@5 concerning the language-specific relevance. 

We make our source code and data publicly available to facilitate reproducibility of the results and their reuse by the research community\footnote{Code: \url{https://github.com/saraabdollahi/LaSER}, \\ Data: \url{https://zenodo.org/record/5735580}}.

\textbf{Contributions.} In summary, our contributions presented in this article are as follows:

\begin{itemize}
    \item We define the new task of language-specific event recommendation. 
    \sg{This task is different from the existing recommendation tasks that focus on the individual user preferences, provide language-independent recommendations, and do not focus on the language-specific relevance and event characteristics.}
    
    \item \sg{We represent the language-specific context through a set of novel features, including spatio-temporal event information, language-specific link data, and publicly available clickstream data that serve as target labels. 
    We blend these features into an architecture for language-specific event recommendations. This architecture relies on the language-specific entity and event embeddings for candidate retrieval and an established learning to rank model.}

    \item We propose novel language-specific embeddings where latent entity representations reflect their neighborhoods and relations in a language-specific knowledge graph and demonstrate that they are beneficial for the candidate generation. 
    
    \item We conduct extensive experiments on real-world data and a user study and demonstrate that our approach outperforms state-of-the-art recommendation methods.
\end{itemize}

The remainder of this article is structured as follows: First, we define the task of language-specific event recommendation in Section \ref{sec:problem}. We present our proposed approach in Section \ref{sec:approach} and introduce datasets used as background knowledge in Section \ref{sec:data}.  
Following that, in Section \ref{sec:eval-setup} we describe our evaluation aims and setup.
Sections \ref{sec:eval_results}, \ref{sec:user_study} and \ref{sec:anecdotal_results} present the results of the ranking evaluation and a user study and discuss anecdotal results and application scenarios.
Section \ref{sec:related_work} provides an overview of related work. Finally, we provide a conclusion in Section \ref{sec:conclusion}.

\section{Problem Statement}
\label{sec:problem}

\ed{
This section defines the notions of a language-specific knowledge graph,
entities, events, and the task of language-specific event recommendation addressed in this article. 
}

\ed{To facilitate recommendation, we introduce a \lkg{}, which models entities, events, and relations in a language context\footnote{Note that in this work we follow a language-specific view, i.e., we do not further distinguish between different sub-communities speaking the same language (e.g., the different English-speaking sub-communities).}.}


\begin{definition} 
\label{def:language_aware_kg}
A \textbf{\lkg{}} is a directed graph $G=(E, R, L)$ whose nodes $E$ represent a set of real-world entities (e.g., persons, places and events), connected via edges $R \subset E \times E$. $L$ is a set of languages.
\end{definition}

In the context of language-specific recommendations, relevant
spatio-temporal features are locations and dates associated with entities.

\begin{definition}
An entity $e \in E$ can be assigned a start and end time $[e.t_s, e.t_e]$ as well as a set of coordinate pairs $e.C$, where each coordinate pair $c \in e.C$ consists of latitude and longitude: $c = (lat,lon), lat \in \mathbb{R}, lon \in \mathbb{R}$.
\end{definition}

For example, the \textit{Summer Olympics 2012} happened from July 27 to August 12, 2012, and are assigned multiple coordinate pairs reflecting different sports venues in London. The entity representing \textit{Winston Churchill} is assigned his birth and death dates (November 30, 1874, to January 24, 1965) and a set of coordinate pairs referring to essential places in his life (e.g., of the Blenheim Palace, his birthplace).

In the context of the language-specific knowledge graph, events are a subset of entities. 
Whereas many definitions of an event exist in the literature, in this work, we follow an event definition by J. Allan et al. proposed in the context of the event detection and tracking within news stories~\cite{allan1998line}:

\begin{definition}
An event $v \in \mathcal{V} \subset E$ is something that happened at a particular time and place.
\end{definition}

Examples of events are the \textit{Summer Olympics 2012}, the \textit{fire at the Notre Dame} in 2020 and the \textit{Coronavirus pandemic in Germany} starting in 2020. For ongoing events like the \textit{Coronavirus pandemic}, the end date is not yet known.

Having introduced the entities, events, and their relations, we can now define the task of language-specific event recommendation. 

In this article, given an entity of user interest referred to as a query entity, we address the new task of recommending relevant events for this entity in a specific language context.
\sg{Note that a query entity can represent a real-world entity or an event.}

\begin{definition}
\label{def:laser}
Given a query entity $e \in E$, a language $l$ and the \lkg{} $G=(E, R, L)$, the task of \textbf{language-specific event recommendation} is to create a ranked list $S_{e,l} = \langle v_1, \dots v_n \rangle$ of events ($v_i \in \mathcal{V}, i \in {1,\dots,n}$). The events in $S_{e,l}$ are sorted in descending order regarding their relevance to the query entity $e$ for the audience speaking the language $l$. 
\end{definition}

For example, consider the recommendation example in Table \ref{tab:intro_example} created from the click counts on Wikipedia articles in specific Wikipedia language editions in April 2021. For the query entity \textit{Coronavirus pandemic} ($e$) and the German language ($l$), this method returns a list $S_{e,l}$ of recommended events $\langle$ \textit{COVID-19 pandemic in Germany}, \textit{COVID-19 recession}, \textit{COVID-19 pandemic in Austria}$\rangle$. Language-specific event recommendation generates a ranked list of events. The query entity may be any node in the \lkg{}.

\section{The \approach{} Approach}
\label{sec:approach}

\begin{figure*}[ht]
    \includegraphics[width=\textwidth]{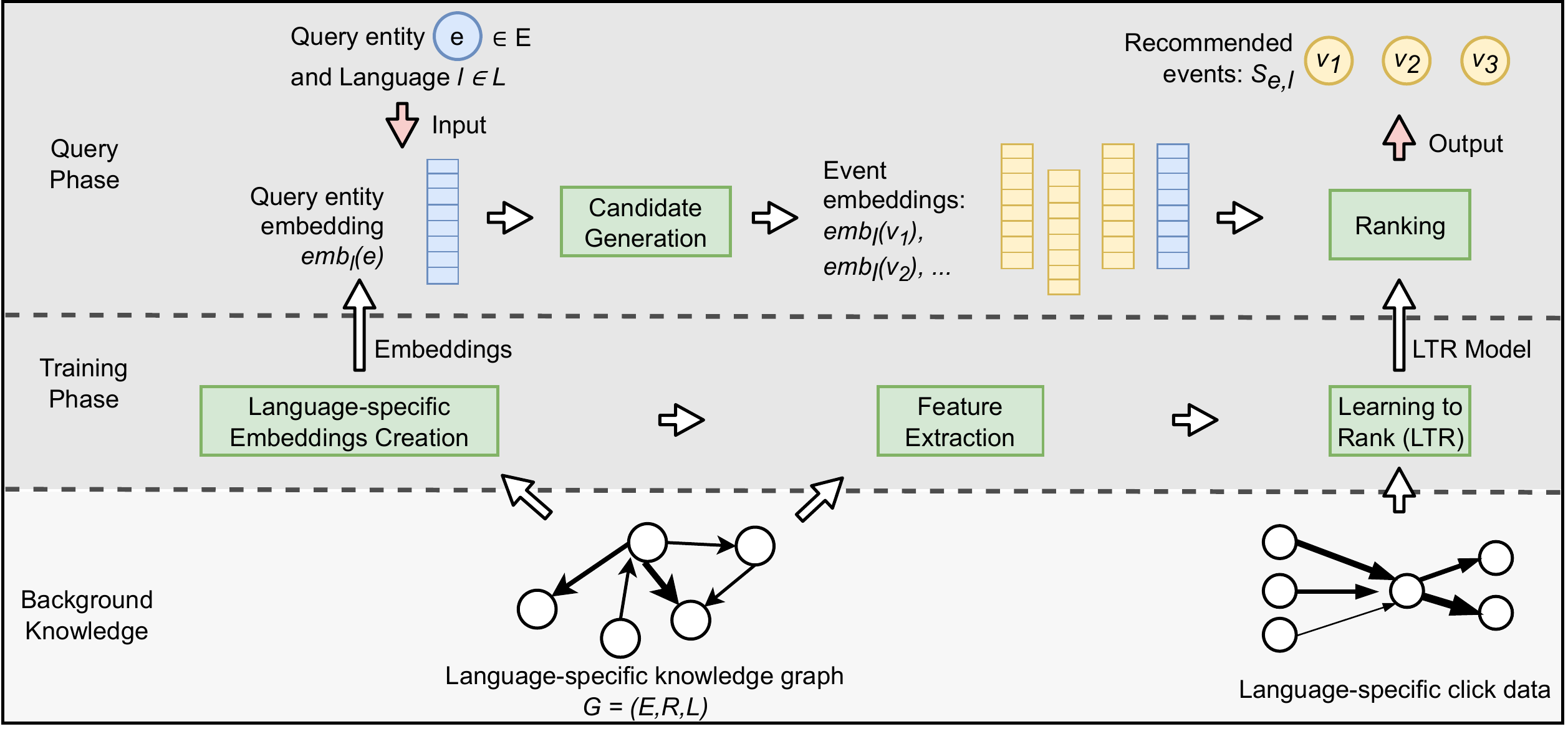}
    \caption{
    The \approach{} overview includes three parts. (i) The background knowledge includes the \lkg{} and the \cg{}. \ed{(ii) In the training pre-processing phase}, the language-specific embeddings and the LTR event ranking model are trained based on this background knowledge. (iii) In the query phase, given a query entity $e$ (e.g., \textit{Coronavirus pandemic}) and a language $l$ (e.g., German) as an input, the embeddings and the LTR ranking model are utilized to generate a language-specific ranked list of events $S_{e,l}$ (e.g., $\langle$\textit{COVID-19 pandemic in Germany}, \textit{COVID-19 recession}, \textit{COVID-19 pandemic in Austria}$\rangle$).
    }\label{fig:overview}
\end{figure*}

In this article, we present \approach{}, a new method for language-specific event recommendation. Figure~\ref{fig:overview} provides an overview of the \approach{} components. \approach{} consists of a training and a query phase. 
These phases rely on background knowledge that includes the \lkg{} and \cg{}.

In the pre-processing training phase, we first create language-specific embeddings based on the \lkg{}. In addition, we train a learning to rank model that learns from \cg{}. This model uses feature values extracted from the \lkg{}, i.e., event characteristics, as well as the relationships between events and entities.

In the query phase, given an input query entity $e \in E$ and a language $l \in L$, we use the embeddings and the trained LTR model to generate a ranked list of events.

\begin{figure*}[ht]
\centering
    \includegraphics[width=0.8\textwidth]{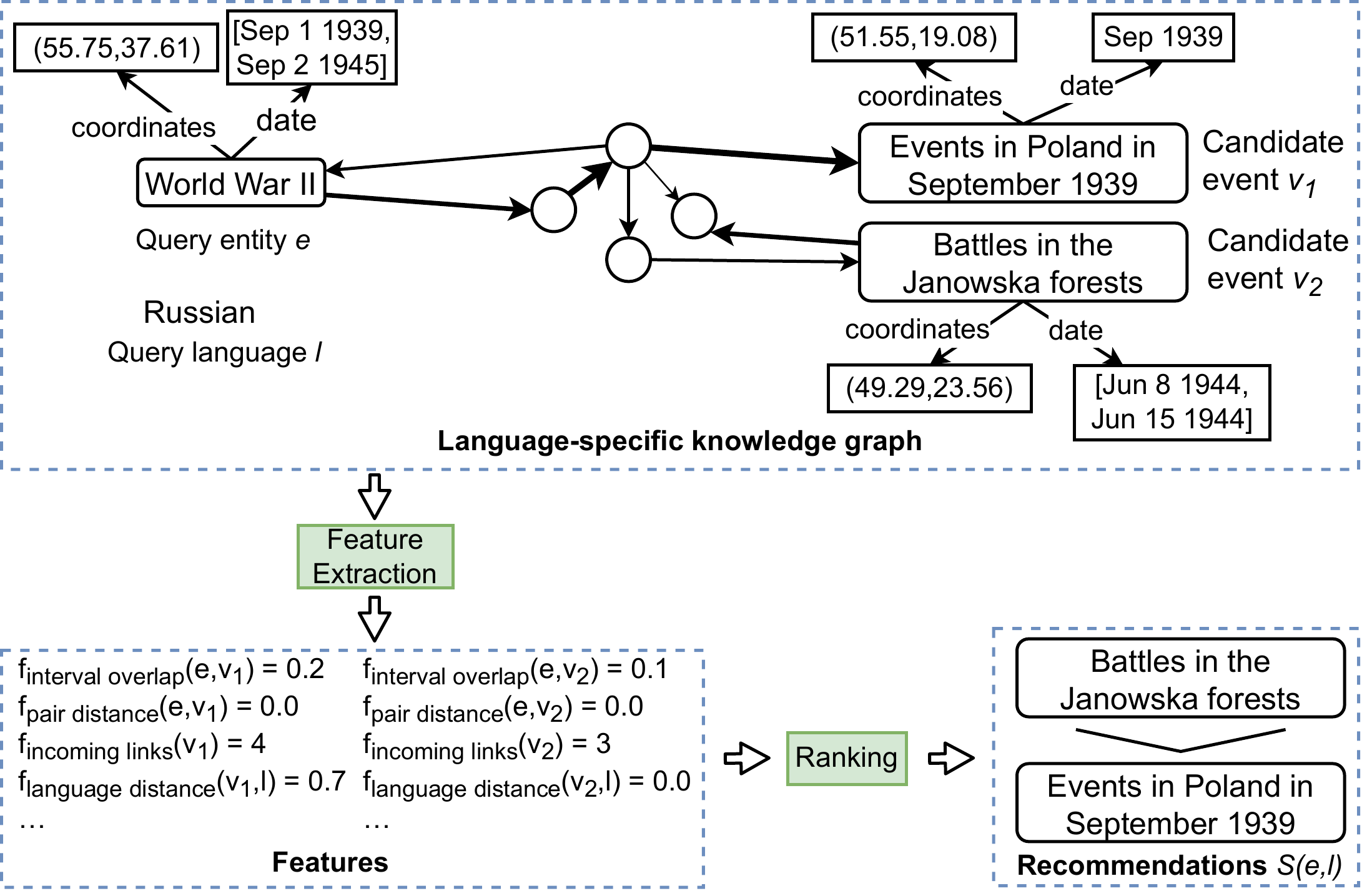}
    \caption{
    Example query looking for events relevant to \textit{World War II} from the Russian perspective. We illustrate two candidate events with selected features for brevity. \approach{} retrieves the candidates, extracts feature values, and ranks the candidates.
    }
    \label{fig:approach_example}
\end{figure*}

\ed{Figure~\ref{fig:approach_example} provides a concrete example of how a query is processed, with a specific focus on the feature extraction step. Here, World War II is the query entity $e$, with Russian as the query language $l$. From the candidate generation, we obtain two events\footnote{We only show two candidate events for brevity. More candidate events can be generated.}: ``Events in Poland in September 1939'' and ``Battles in the Janowska forests''. 
From the links and the spatio-temporal information in the language-specific knowledge graph, \approach{} extracts feature values concerning the query entity, the candidate events and the language. Finally, \approach{} ranks the candidate events in order of their language-specific relevance.}

In this section, we describe the background knowledge and training and query phases of \approach{} in more detail.

\subsection{Background Knowledge of \approach{}}
\label{subsec:background_knowledge}

The \approach{} approach relies on background knowledge, including the \lkg{} and \cg{}.

\subsubsection{Language-specific Knowledge Graph}
\label{subsec:lkg}

Following Definition~\ref{def:language_aware_kg}, the \lkg{} $G=\allowbreak (E,\allowbreak R,\allowbreak L)$ represents entities, their spatial and temporal characteristics, and relationships in the context of a specific language $l$.

\subsubsection{Language-specific Click Data}
\label{subsec:cd}

The \cg{} provides training labels for the ranking model. 
This data is extracted from the Wikipedia Clickstream and represents real user interactions with Wikipedia articles corresponding to the entities in the \lkg{}. From such user interactions, we infer the language-specific relevance scores for an entity $e \in E$ and an event $v \in \mathcal{V}$: $rel_{interaction}(e,v,l) \in [0,1]$. Such values are derived by normalizing click counts in the Wikipedia Clickstream of a language $l \in L$ regarding click counts in all languages $L$. That way, the $rel_{interaction}$ scores reflect the language-specific relevance. We provide more details regarding this normalization in Section~\ref{sec:data}.

\subsection{Training Phase}
\label{subsec:training_phase}

The goal of the training phase is to create language-specific embeddings and train an event ranking model. 
\ed{The training phase is conducted as a pre-processing and does not impact the query efficiency.}
This phase consists of the following three steps:
\begin{enumerate}
\item Language-specific embeddings creation: From the \lkg{}, we create language-specific embeddings of entities and events.
\item Feature extraction: For a pair of an entity and an event, we extract feature values representing different characteristics of the event and the pair. Example features are the event popularity, the spatial distance between the entity and the event and their embedding similarity.
\item Learning to rank: We incorporate the features to train an LTR model which ranks events regarding their relevance to the query entity.
\end{enumerate}
In the following, we describe these steps in more detail.

\subsubsection{Language-specific DeepWalk Embeddings}
\label{ec:embeddings_creation}

To leverage the information of entities and the structure of the \lkg{} efficiently, we propose a language-specific embedding technique that learns continuous vector representation of entities representing their relations in a language $l$. This technique maps the entities to low-dimensional vectors, which are similar in cases where two entities appear close to each other in the language-specific context of $l$.

To create language-specific embeddings, we utilize DeepWalk~\cite{perozzi2014deepwalk}, and follow a uniform random sampling approach: the next entity to visit in the random walk is chosen uniformly from all neighbors of the current entity.

After creating a set of random walks following the DeepWalk approach, we train a Word2Vec model. The resulting language-specific embeddings are utilized for (i) creating a candidate set of events relevant to the query entity and (ii) for measuring the language-specific relevance between the query entity and the event.

\subsubsection{Feature Extraction}
\label{subsubsec:feature_extraction}

To model event relevance to the query entity $e$ in the context of language $l$, we extract $10$ features from the event $v$ and the entity $e$. This set of features $F$ includes four groups, covering different entity aspects: spatial, temporal, link-based, and embedding-based features. Some of these features are computed for the event and the language only: $f(v, l) \mapsto \mathbb{R}, v \in \mathcal{V}, l \in L$. Other features are computed for the pair of a query entity and an event in the specific language: $f(v, l, e) \mapsto \mathbb{R}, v \in \mathcal{V}, l \in L, e \in E$ (language-dependent features) or irrespective of the language: $f(v, e) \mapsto \mathbb{R}, v \in \mathcal{V} , e \in E$.

\textbf{Spatial Features ($Features_{spatial}$)}: As per challenge (C1), events have a spatio-temporal dimension. Consequently, spatial features are used to capture spatial dependencies between an event and the language, as well as between an event and the query entity. 

\begin{itemize}
\item Language distance: This continuous feature denotes the spatial distance between the event and the set of countries $\mathfrak{C}_l=\left \{ \mathfrak{c}_1, \mathfrak{c}_2, \ldots, \mathfrak{c}_i \right \}$ where $l$ is an official language\footnote{We extract the set of countries where $l$ is an official language from Wikidata using the \textit{official language} property (\url{https://www.wikidata.org/wiki/Property:P37}). For example, we extract the countries Germany, Switzerland, Austria, and Liechtenstein for the German language.}: 
   
\begin{equation}
       f_{\textnormal{language distance}}(v,l) = \min_{c \in v.C, \mathfrak{c} \in \mathfrak{C}_l} \textnormal{distance}(c, \mathfrak{c}).
       \label{eq:language_distance}
\end{equation}
   
An event can be assigned multiple coordinate pairs ($v.C$), and there can be multiple countries where $l$ is an official language. This feature is based on the closest combination of event coordinate pairs and a country. 
$distance$ represents the distance between a point and a polygon (i.e., a country) in kilometers (or $0$, if the point is located inside the polygon).

The intuition behind the language distance feature is that the distance can directly impact the interest of the language audience. For example, due to the location, the \textit{2004 Summer Olympics} held in Athens are expected to be more relevant to the Greek language community than the \textit{2008 Summer Olympics} located in Beijing.
   
\item Pair distance: This continuous feature represents the spatial distance between the query entity $e$ and the event $v$: 

   \begin{equation}
       f_{\textnormal{pair distance}}(v,e) = \min_{c_1 \in v.C, c_2 \in e.C} \textnormal{distance}(c_1, c_2).
   \end{equation}
   
   This feature considers the minimum distance between any of the coordinate pairs of $v$ and $e$. In contrast to the language distance in Equation~\ref{eq:language_distance}, here $distance$ represents the spatial distance between two points. We assume that the lower their distance, the more relevant the event is to the query entity.
\end{itemize}

\textbf{Temporal Features ($Features_{temporal}$)}: To take the temporal closeness of the query entity $e$ and an event $v$ into account, we employ temporal features. 

\sg{The intuition behind the temporal features is that an event is expected to be more relevant regarding the query entity if they happened simultaneously. The extent of such temporal coincidence is computed through two different features which measure both the temporal overlap and the distance.}

\begin{itemize}
\item Interval overlap: This feature indicates the overlap between the time intervals of the query entity $e$ and an event $v$:

\begin{equation}
\begin{split}
& f_{\textnormal{interval overlap}}(v,e) \\
& \text{ } = \begin{cases}
0, \text{ if }v.t_s > e.t_e\text{ or }e.t_s > v.t_e \\
| [max(v.t_s, e.t_s), min(v.t_e, e.t_e) ] |, \text{ else}.
\end{cases}
\end{split}
\end{equation}
    
\item Begin time distance: The start time distance feature represents the time difference between the start times of the query entity $e$ and an event $v$:

\begin{equation}
f_{\textnormal{begin time distance}}(v,e) = |v.t_s - e.t_s|.
\end{equation}
    
\end{itemize}

\sa{Temporal features represent overlap and distance based on the number of days and have discrete values.}

\textbf{Link-based Features ($Features_{links}$)}: 
The features in this category represent overall event importance and the similarity of the query entity $e$ and an event $v$ based on their \lkg{} neighborhoods. 
We assume that an event is more relevant to an entity if they appear in the same contexts, i.e., have a similar neighborhood in the graph. To measure such similarity of neighborhoods, we consider the number of incoming and outgoing links as well as the shared links between them. 
We obtain the link counts from the specific Wikipedia language editions.
Given a set of links $W_l = E \times E$ in a language-specific Wikipedia edition in a language $l \in L$, there is a link from one entity $e \in E$ to another entity $e_n \in E$, if $(e,e_n) \in W_l$. 
%
%
To compute the overall importance of the event, we use the number of incoming links and outgoing links of $v$. To measure the similarity of the neighborhoods, we consider the number of shared incoming and outgoing links.

\begin{itemize}
\item Number of incoming links: We estimate an overall importance of an event in a language context based on its link count in the Wikipedia link set $W_l$ of the language $l$:

\begin{equation}
f_{\textnormal{incoming links}}(v,l) = | \{ (e_i,v) \in W_l \} |.
\end{equation}

\item Number of outgoing links: In analogy to the number of incoming links, we also consider the number of outgoing links. This feature represents the general interaction of the event with other entities in $W_l$:
   
\begin{equation}
f_{\textnormal{outgoing links}}(v,l) = | \{ (v,e_i) \in W_l \} |.
\end{equation}

\item Number of shared incoming links: We estimate the similarity between the query entity $e$ and an event $v$ in terms of their interlinking with their neighbors in a specific language. A shared incoming link represents the situation where an entity $x \in E$ refers to both $v$ and $e$ in $W_l$:

\begin{equation}
f_{\textnormal{shared incoming links}}(v,l,e) =  \{ x | (x, v) \in W_l \land (x, e) \in W_l\}.
\end{equation}

\item Number of shared outgoing links: We also consider shared outgoing links. This number represents the similarity between the query entity $e$ and an event $v$ in terms of their interaction with other entities in $W_l$. A shared outgoing link represents the situation where $v$ and $e$ refer to an entity $x \in E$ in the context of language $l$:

\begin{equation}
f_{\textnormal{shared outgoing links}}(v,l,e) = \{ x | (v,x) \in W_l \land (e,x) \in W_l\}.
\end{equation}
\end{itemize}
\sa{All link-based features have discrete values.}

In addition, we use the Milne-Witten relatedness score that is often used to estimate the semantic relatedness between Wikipedia articles~\cite{witten2008effective}.

\begin{itemize}

\item Milne-Witten relatedness: This feature value is computed using the Wikipedia link-based measure proposed by Milne and Witten~\cite{witten2008effective}.

\begin{equation}
\begin{split}
{} & f_{\textnormal{Milne-Witten}}(e,v,l) =  \\
& 1-\frac{log(max(|In_{e}|,|In_{v}|))-log(|In_{e}\cap In_{v}|)}{log(|E|)-log(min(|In_{e},In_{v}|))},
\end{split}
\end{equation}

where $In_{e} = \{ e | (e, e_i) \in W_l, e \in E \}$ and $In_{v} = \{ v | (v, v_i) \in W_l, v \in \mathcal{V} \}$ are the sets of all incoming links to the query entity $e$ and an event $v$ in the Wikipedia link set $W_l$, respectively. The continuous Milne-Witten relatedness feature is bound between $0$ and $1$.
\end{itemize}

\textbf{Embedding-based Features ($Features_{embeddings}$)}: The embedding-based features make use of the previously computed language-specific embeddings.

\begin{itemize}
    \item Embedding similarity: We compute the cosine similarity between the language-specific embeddings of the query entity $e$ and an event $v$:
    
    \begin{equation}
f_{\textnormal{embedding similarity}}(v,l,e) = cos(emb_l(v), emb_l(e)).
\label{eq:embsim}
\end{equation}

This continuous feature is bound between $0$ and $1$. We assume that $v$ is relevant to $e$ if their embedding vectors are close in the embedding space, as reflected by their cosine similarity.

\end{itemize}

\subsubsection{Learning to Rank}
\label{subsubsec:letor}

To rank the events relevant to the query entity, we train a learning to rank model that takes feature values as an input and is trained to predict the ranking inferred from the \cg{}. 
In the context of the LTR model, the problem of language-specific event recommendation is defined as follows: 
Given a training set of language-specific relevance values between entities and events as well as their features, learn a scoring function that approximates the language-specific relevance $rel_{interaction}(e,v,l)$ for the query entity $e$ and an event $v$ in a language $l$.

We train a tree ensemble model to learn an optimal ranking of the language-specific relevance scores using LambdaMART \cite{burges2010ranknet}. LambdaMART is an LTR algorithm that uses gradient boosted decision trees with a cross-entropy cost function. 
In the literature, LambdaMART has been shown to outperform neural ranking models in information retrieval tasks \cite{guo2020deep}. Using LambdaMART, we perform a list-wise ranking where the normalized discounted cumulative gain (nDCG) is maximized.

\subsection{Query Phase}

In the query phase, \approach{} takes the query entity $e \in E$ and a language $l \in L$ given by the user as input and recommends a language-specific ranking of events $S_{e,l}$ as an output\footnote{We assume that users can select a query entity from the \lkg{}, e.g. via its label.}.

The query phase consists of the following two steps:

\begin{enumerate}
\item Candidate generation: A set of candidate events is generated based on the language-specific embeddings.  
\item Ranking: The candidate events are ranked by the previously trained LTR model.
\end{enumerate}

\subsubsection{Candidate Generation}
\label{sec:candidate_generation}

Due to numerous events in the \lkg{}, it is not feasible to compute the relevance scores of all events and to rank them. Therefore, we collect a set of candidate events that are likely to be among the recommended events for the query entity. 
More specifically, similar to the idea of \cite{ni2020layered}, we select $k$ events which are most similar to the query entity $e$ based on the embedding similarity 
computed using Equation \ref{eq:embsim}.
Such a candidate set can be obtained efficiently and reflects structural similarities. 

\subsubsection{Ranking}
Finally, for each candidate event $v$, we compute its feature values as well as the feature values between the query entity $e$ and $v$ based on the \lkg{}. Given the input set of all candidate events and their feature values, we employ the LTR model trained in the training phase to estimate the language-specific relevance scores. The resulting scores are then used to sort the candidate events according to their relevance and create the set of recommendations $S_{e,l}$.

\section{Extraction of Background Knowledge}
\label{sec:data}

\approach{} requires a \lkg{} and \cg{} described in Section~\ref{subsec:background_knowledge} as background knowledge. In the following, we describe the extraction of both datasets in more detail.

\subsection{\lkgCap{}}

Entities and their attributes (start and end time and coordinate pairs) in the \lkg{} are collected from the EventKG knowledge graph \cite{gottschalk2019eventkg}. 
%
\ed{
Note that although we are interested in the language-specific information, and thus conceptually speak about a language-specific knowledge graph, from the practical perspective we can also extract such information from multilingual sources, such as EventKG, directly.
}

\subsection{Entities from EventKG}
\label{sec:entities-eventkg}

EventKG \cite{gottschalk2019eventkg} is a multilingual knowledge graph that contains semantic information regarding events, their relations and temporal information about real-world entities. Such information builds the basis for the \lkg{} $G=(E, R, L)$ defined in Definition~\ref{def:language_aware_kg}. Specifically, $E$ represents the entities of EventKG.

\begin{figure*}
  \begin{subfigure}{0.31\textwidth}
    \includegraphics[width=\linewidth]{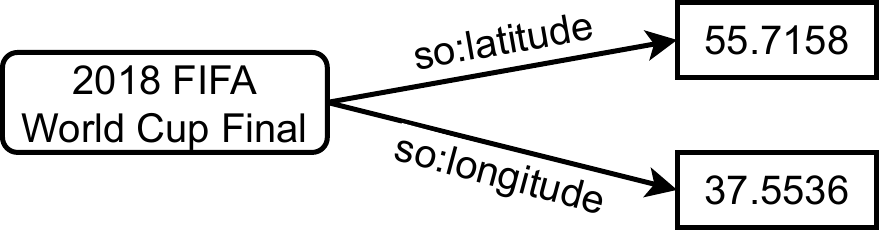}
    \caption{Event with coordinates.}
    \label{fig:coordinates_1}
  \end{subfigure}%
  \hspace*{\fill}
  \begin{subfigure}{0.32\textwidth}
    \includegraphics[width=\linewidth]{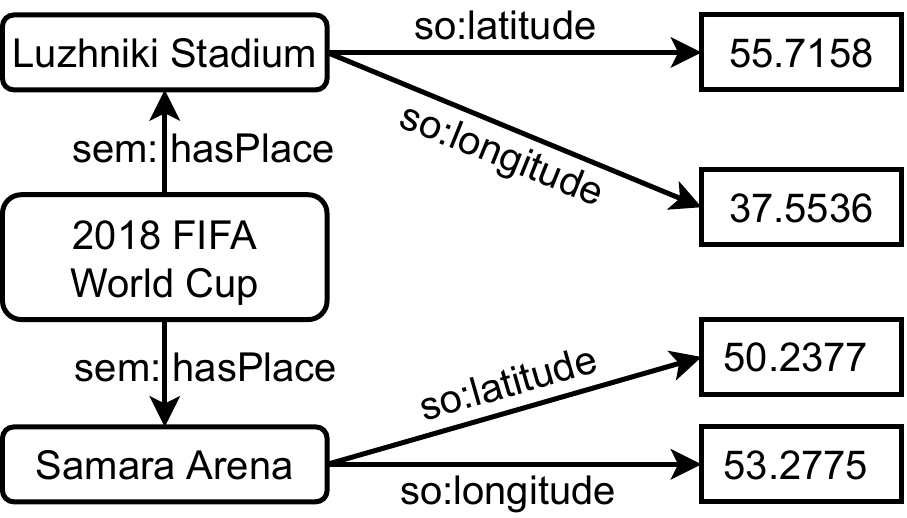}
    \caption{Event with locations.}
    \label{fig:coordinates_2}
  \end{subfigure}
  \hspace*{\fill}
  \begin{subfigure}{0.32\textwidth}
    \includegraphics[width=\linewidth]{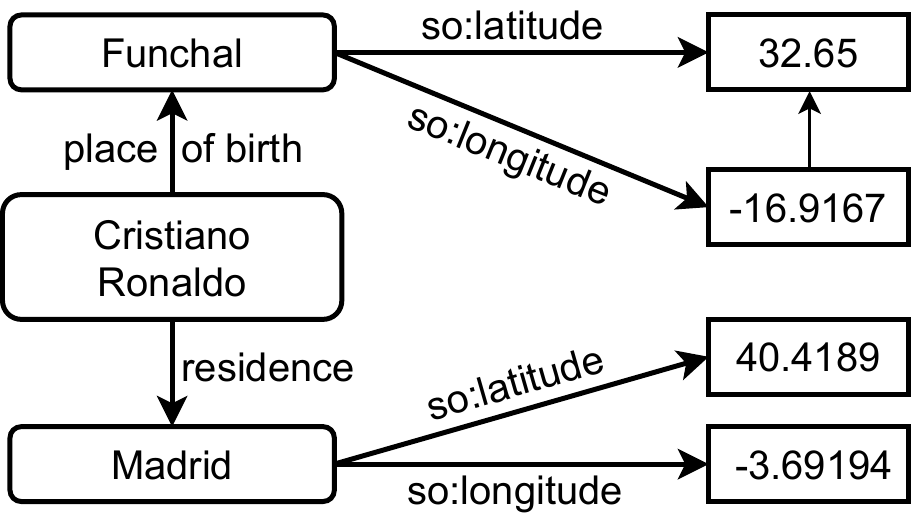}
    \caption{Entity with relations to locations.}
    \label{fig:coordinates_3}
  \end{subfigure}
\caption{Three examples of coordinate pairs for selected entities. For readability, we use entity and property labels instead of their URIs except for few selected properties.} \label{fig:coordinates}
\end{figure*}

The EventKG entities typed as \voc{sem}{Event}\footnote{\schema{sem}: \url{http://semanticweb.cs.vu.nl/2009/11/sem/}, Simple Event Model~\cite{van2011design}} make up the event set $\mathcal{V} \subset E$.
 To retrieve the start and end times of entities in $E$, we use EventKG's \voc{sem}{has\-Begin\-Time\-Stamp} and \voc{sem}{has\-End\-Time\-Stamp} relations. The set of coordinate pairs for each entity is collected from a set of relations in EventKG. As shown in Figure~\ref{fig:coordinates}, there are different options to retrieve the coordinates from the \voc{so}{latitude}\footnote{\schema{so}: \url{http://schema.org/}} and \voc{so}{longitude} triples. For events and locations, coordinates are often directly assigned to them (Figure~\ref{fig:coordinates_1}). Some events are assigned coordinates via \voc{sem}{hasPlace} (Figure~\ref{fig:coordinates_2}). Finally, the entities can be connected to locations via other properties (Figure~\ref{fig:coordinates_3}). For each entity, we select coordinates from one of these options where coordinates are available.
\sa{Following this process, $40\%$ of entities have temporal and $48\%$ spatial information. In terms of events, the numbers are $82\%$ and $81\%$ for temporal and spatial features, respectively.}

\subsection{Link Counts from Wikipedia}
To compute the link features, we use Wikipedia. Wikipedia is a multilingual encyclopedia available in more than 300 languages\footnote{\url{https://en.wikipedia.org/wiki/List\_of\_Wikipedias}}. 
Wikipedia is actively edited by volunteers worldwide, and thus reflects the cultural preferences of audiences in different language communities \cite{hecht2010tower,rogers2013digital}. Each Wikipedia language edition covers a different set of interlinked articles. To profit from these language-specific link structures that potentially represent different linguistic points of view and language-specific information asymmetries, we use Wikipedia as an underlying resource to compute the link features. 
We extract the Wikipedia links $W_l$ from the German, French and Russian Wikipedia language editions in our experiments\footnote{The German, French and Russian Wikipedia are among the top-6 most actively edited Wikipedia editions. English, the most active Wikipedia edition, is often edited by users of other languages and thus has a less clear language-specific focus~\cite{hale2014multilinguals}.}.
Table~\ref{tab:wikipedia_statistics} provides statistics of the extracted Wikipedia links.

\begin{table}[h]
\center
\caption{Statistics of the links extracted from three Wikipedia language editions.}
\label{tab:wikipedia_statistics}
\begin{tabular}{@{}lrrr@{}}
\toprule
 & \multicolumn{1}{c}{\textbf{German}} & \multicolumn{1}{c}{\textbf{French}} & \multicolumn{1}{c}{\textbf{Russian}} \\ \midrule
\textbf{Entities (Nodes)} & 3,028,223 & 2,259,750 & 2,242,357 \\
\textbf{ thereof Events} & 87,573 & 89,130 & 59,557 \\
\textbf{Links (Edges)} & 51,001,819 & 41,526,761 & 29,106,336 \\ \bottomrule
\end{tabular}
\end{table}

\subsection{\cgCap{}}
\label{subsubsec:eventkgclick}

To train the LTR model for event recommendation, \approach{} requires click data reflecting the language-specific user preferences. We adopt the \ekgc{} dataset created in our previous work~\cite{abdollahi2020eventkg+} to obtain such information. \ekgc{} is a publicly available, cross-lingual dataset that reflects the language-specific relevance of events and their relations. This way, we facilitate the reproducibility of results addressing the challenge (C2).

\ekgc{} was created on top of two data sources: (i) the EventKG knowledge graph, which provides the set of events and (ii) the Wikipedia Clickstream dataset~\cite{clickstream} that reflects how users explore articles in different language editions. Precisely, the Wikipedia Clickstream contains counts of how often users clicked on a specific link in a Wikipedia article. For example, the Wikipedia Clickstream contains the following information: in 2020, the links leading to the article \textit{2020 United States presidential election} on the Wikipedia article about \textit{Bernie Sanders} were clicked $2,196$ times in the Russian Wikipedia. Another example was shown in Table~\ref{tab:intro_example}. Table~\ref{tab:Clickstream_statistics} provides statistics extracted from the Wikipedia Clickstream in 2020.

\begin{table}[h]
\center
\caption{Statistics of the Wikipedia Clickstream in 2020 in three languages. Click pairs are all recorded clicks between two Wikipedia articles in the respective language and time span.}
\label{tab:Clickstream_statistics}
\begin{tabular}{@{}lrrr@{}}
\toprule
 & \multicolumn{1}{c}{\textbf{German}} & \multicolumn{1}{c}{\textbf{French}} & \multicolumn{1}{c}{\textbf{Russian}} \\ \midrule
\textbf{Entities} & $1,222,070$ & $994,381$ & $831,703$ \\
\textbf{Clicked events} & $40,223$ & $46,557$ & $33,712$ \\
\textbf{Click pairs} & $6,862,960$ & $5,192,491$ & $5,264,813$ \\ \bottomrule
\end{tabular}
\end{table}

While the Wikipedia Clickstream provides information regarding \sa{entities and their}
relations for the specific Wikipedia language editions, the Clickstream does not put the different language editions in relation to each other. Thus, the clickstream does not fully reflect the language-specific relevance. 
In this work, we extend the \ekgc{} dataset to use it as a training corpus for the proposed \approach{} approach as described in the following.

To compute language-specific relevance scores in \ekgc{}, we follow our previous work~\cite{abdollahi2020eventkg+}. The idea behind the language-specific relevance scores is to take all languages available in the Wikipedia Clickstream ($\mathfrak{L}$) into account and normalize the click counts regarding the number of clicks in other languages. For example, while the articles regarding  \textit{Coronavirus pandemic} and \textit{2021 German federal election} are often clicked in the German Wikipedia,  \textit{German federal election} has a higher language-specific relevance score. This score is higher because the relative number of clicks on  \textit{German federal election} is higher in the German Wikipedia than in any other Wikipedia language edition, highlighting the language-specific relevance of the event in the German context.

Formally, given a set of click counts from source to target entities ($clicks$), we first create \textit{balanced click counts} between a source entity $e_s \in E$ and a target entity $e_t \in E$ as follows: 

\begin{equation}
\begin{multlined}
balanced\_clicks(e_s,e_t,l) = \\
clicks(e_s,e_t,l) \cdot \frac{\sum_{l' \in \mathfrak{L}}\sum_{e_s' \in E}\sum_{e_t' \in E} clicks(e_s',e_t',l')}{\sum_{e_s' \in E}\sum_{e_t' \in E} clicks(e_s',e_t',l)}.
\end{multlined}
\end{equation}

The language-specific relevance value between an entity $e \in E$ and an event $v \in \mathcal{V}$ introduced in Section~\ref{subsubsec:letor} is then computed as follows:

\begin{equation}
\begin{multlined}
rel_{interaction}(e,v,l) = \\
\frac{balanced\_clicks(e,v,l)}{\sum_{l' \in \mathfrak{L}} balanced\_clicks(e,v,l')} \in [0,1].
\end{multlined}
\end{equation}

Following this procedure, we created an extended version of \ekgc{} covering the whole year of 2020 in the languages German, French and Russian\footnote{In comparison to the first version of \ekgc{}, the new version covers the entire year 2020 and considers all available languages ($\mathfrak{L}$) when computing the relevance scores.}. We make the extended version of \ekgc{} publicly available\footnote{\url{https://github.com/saraabdollahi/EventKG-Click}}. The statistics of this dataset are presented in Table~\ref{tab:eventkgclick_statistics}. 

\begin{table}[h]
\center
\caption{Statistics of the \cg{} obtained from \ekgc{}.}
\label{tab:eventkgclick_statistics}
\begin{tabular}{@{}lrrr@{}}
\toprule
 & \multicolumn{1}{c}{\textbf{German}} & \multicolumn{1}{c}{\textbf{French}} & \multicolumn{1}{c}{\textbf{Russian}} \\ \midrule
\textbf{Source Entities} & $117,281$ & $104,331$ & $97,212$ \\
\textbf{Events} & $40,223$ & $46,557$ & $33,712$ \\
\textbf{Relevance Pairs} & $304,564$ & $271,243$ & $254,910$ \\
\bottomrule
\end{tabular}
\end{table}

\section{Evaluation Aims and Setup}
\label{sec:eval-setup}

The evaluation aims to assess the performance of the main components of \approach{}. 
First, we aim to assess the effectiveness of 
the proposed language-specific embedding method and its impact on the candidate generation step of \approach{}. 
Second, we conduct an evaluation of the recommendations, where we assess \approach{}'s results using the relevance labels obtained from the EventKG+Click dataset as a ground truth. Third, we analyze the effectiveness of the features adopted by the proposed \approach{} approach. 
Fourth, we conduct a user study to assess the recommendation quality from the user perspective. 
Finally, we manually analyze anecdotal evaluation results and discuss potential application scenarios of the proposed \approach{} approach.

This section introduces the ground truth for the recommendation evaluation, presents the embedding and recommendation baselines, describes the evaluation metrics, and provides the implementation details. 

\subsection{Ground Truth Creation}
\label{subsec:ground_truth}

To train \approach{} and evaluate the language-specific recommendation, we created a ground truth of language-specific event recommendations from EventKG+Click (see Section~\ref{subsubsec:eventkgclick}). For each considered language $l$, this ground truth $GT_l$ contains query entities together with a ranked list of events and is composed as follows:

\begin{equation}
\begin{split}
GT_l = & \{(e, \langle v_1, \dots, v_n, v^-_1, \dots, v^-_n\rangle ) \\
& | rel_{interaction}(e,v_i,l) \geq rel_{interaction}(e,v_j,l) \\
& \forall 1 \leq i < j \leq n\},
\end{split}
\end{equation}

where $v^-_i$ denote negative examples, i.e., randomly chosen events not related to the query entity: $rel_{interaction}(e,v^-_i,l)=0$. In other words, we select all entities in EventKG+Click for which events are provided and rank these events according to their $rel_{interaction}$ score. Each ranked event list is extended with randomly chosen negative examples of the same number as positive events.

We make the ground truth available online\footnote{\url{https://zenodo.org/record/5735580}}.

\subsection{Embedding Methods}
\label{sec:embeddings}

\sg{\approach{} relies on node embeddings both for candidate generation and as a feature for ranking. We compare the following embedding methods in our evaluation.}

\subsubsection{DeepWalk Embedding}
As described in Section~\ref{ec:embeddings_creation}, DeepWalk~\cite{perozzi2014deepwalk} is an embedding approach that learns latent representations of nodes in a network and generalizes neural language models to process sets of randomly generated walks in analogy to sentence-based text embedding models. We compute the DeepWalk embeddings from the language-specific Wikipedia links in each language $l$. 

\subsubsection{Node2Vec Embedding}
Node2Vec~\cite{grover2016node2vec} is an embedding approach that learns continuous feature representations for nodes in a network that maximizes the likelihood of preserving a network
neighborhood of nodes. The authors of Node2Vec designed a biased random walk procedure, which efficiently explores diverse neighborhoods. Unlike DeepWalk~\cite{perozzi2014deepwalk} that creates embeddings on an unweighted graph using uniform random walks, Node2Vec considers edge weights to conduct a biased random walk. This biased random walk has the flexibility of exploring the network neighborhoods by a trade-off between breadth-first search and depth-first search. 
We set the Node2Vec parameters $p=4$ and $q=0.5$, following the experimental results in \cite{grover2016node2vec}. 
\sa{As edge weights, we take the average of shared incoming links and shared outgoing links from the language-specific Wikipedia links.}

\subsubsection{Wikipedia2Vec Embedding}
Wikipedia2Vec~\cite{yamada2020wikipedia2vec} jointly learns word and entity embeddings by applying the skip-gram model on the Wikipedia link graph, Wikipedia texts and the context terms of Wikipedia links. We use language-specific Wikipedia2vec embeddings pre-trained on the German, French, and Russian Wikipedia\footnote{\url{https://wikipedia2vec.github.io/wikipedia2vec/pretrained/}}.

\subsubsection{TransE Embedding}
TransE~\cite{bordes2013translating} models relationships by interpreting them as translations operating on the low-dimensional entity embeddings. We use TransE embeddings\footnote{\url{https://graphvite.io/docs/latest/pretrained_model.html}} pre-trained on the Wikidata5m dataset, which is not language-specific \cite{wang2021kepler}.

\subsection{Recommendation Baselines}
\label{sec:baselines}

To compare the proposed \approach{} approach to the state-of-the-art recommendation baselines, we need to ensure that the baselines: (i) represent the state-of-the-art in the entity or event recommendation, (ii) can be applied to the novel task of language-specific event recommendation proposed in this work and (iii) are reproducible, i.e., do not depend on any proprietary data. 

\sg{Therefore, following the evaluation procedure in \cite{tran2017beyond}, we evaluate \approach{} against four recommendation baselines, which use publicly available data and have different facets: Milne-Witten, which models the relatedness of entities based on links, the embedding-based methods DeepWalk and Node2Vec, as well as SuperGAT, an attention-based graph neural network.} 
We use each of the recommendation baselines to provide a relevance score between an entity $e\in E$ and an event $v \in \mathcal{V}$ in a language $l$, which is used for the event ranking in the language-specific event recommendation.
In the following, we describe these baselines in more detail.

Based on the considerations above, we exclude methods that focus on highly specialized recommendation aspects, such as \cite{zhang2016probabilistic} due to their temporal focus. We also exclude entity recommendation approaches proposed in \cite{ni2020layered} which depends on the proprietary Yahoo! search logs for feature extraction, and \cite{huang2018learning} which heavily relies on proprietary click-through data and search logs of the Baidu Web search engine. As these datasets are not publicly available, recommendation methods that heavily depend on these datasets cannot be reproduced and compared to our approach. Note that these methods address the general entity recommendation task, as opposed to the language-specific event recommendation we address. In contrast, we select the baselines which can be applied to the \lkg{} to address the language-specific relevance.

\subsubsection{Milne-Witten Recommendation Baseline}
The Milne-Witten score measures the semantic relatedness between two entities based on the Wikipedia hyperlink structure~\cite{witten2008effective}. For this baseline, we rank events based on their feature value $f_\textnormal{Milne-Witten}(e,v,l)$ defined in Section~\ref{subsubsec:feature_extraction}.

\subsubsection{DeepWalk Recommendation Baseline}
As the DeepWalk recommendation baseline, we rank events regarding the cosine similarity between the DeepWalk embeddings 
of the query entity $e$ and an event $v$.

\subsubsection{Node2Vec Recommendation Baseline}
As the Node2Vec recommendation baseline, we rank events regarding the cosine similarity between the Node2Vec embeddings of the query entity $e$ and an event $v$.

\subsubsection{SuperGAT Recommendation Baseline}

We compare \approach{} to SuperGAT, a state-of-the-art self-supervised graph attention network~\cite{kim2021find}. 
The architecture of the SuperGAT recommendation baseline consists of an encoder and a decoder component.
In the encoder component, the nodes in the language-specific knowledge graph are embedded using the SuperGAT network.
In the decoder component, pairs of negative and positive examples are created from the language-aware click data. Each such pair consists of one positive (query entity, event) example ($e,v_1$) and a negative (query entity, event) example ($e,v_2$), such that $rel_{interaction}\allowbreak(e,v_1,l)\allowbreak >\allowbreak rel_{interaction}\allowbreak(e,v_2,l)$ for the given language $l$. Based on such pairs, the objective is then to minimize the loss function.
We adopt the margin ranking loss function typically used in recommender systems~\cite{sun2021hgcf}. 
Here, the goal is to rank positive (query entity, event) examples above their corresponding negative (query entity, event) examples.
We use three hidden layers with an embedding size of $64$ and a learning rate of $0.01$.

\subsection{Evaluation Metrics} 
\label{sec:ev-metrics}

For candidate generation, we report candidate recall, i.e., the fraction of events in the ground truth included in the candidate events. 

To evaluate the recommendation quality, we use the normalized discounted cumulative gain (nDCG@10) and mean average precision (MAP@10). nDCG@10 compares the top-$10$ ranked events against the ideal ranking of the ground truth and rewards relevant events in the higher positions, where the ideal ranking achieves an nDCG@10 score of $1.0$~\cite{jarvelin2002cumulated}. 
MAP@10 averages over the average precision scores (AP@10) of each query entity in the ground truth, where AP@10 is the sum of precision scores (precision@k for $k=\{1,...,10\}$) divided by the total number of relevant events in the top-$10$ ranked results.

For evaluating the user study, we employ mean average precision (MAP@5) regarding the user's relevance judgments. 
When reporting user study results for selected events, we use average precision (AP@5).

\subsection{Implementation Details}
\label{sec:implementation}

\approach{}, DeepWalk, Node2Vec and Milne-Witten are implemented in Python 3.7 and the SuperGAT baseline is implemented using PyTorch Geometric \cite{Fey/Lenssen/2019}. All experiments are conducted on a Linux machine with Intel(R) Xeon(R) Silver 4210 CPU@ 2.20 GHz and 1 TB memory.
We train \approach{} on EventKG+Click in each language separately. 
To train the LTR model, we have used the XGBoost library~\cite{chen2016xgboost} which provides a regularizing gradient boosting framework. 

\section{Evaluation Results}
\label{sec:eval_results}

In this section, we first present the results of the candidate generation and the recommendation evaluation, where we 
assess the performance of \approach{} based on the ground truth obtained from EventKG+Click. 
Then, in a feature analysis, we assess the impact of different feature groups on the \approach{}'s performance. 
The queries considered in this section correspond to the source entities in the EventKG+Click dataset presented in Table \ref{tab:eventkgclick_statistics}.

\subsection{Candidate Generation Evaluation}
\label{subsec:embeddingstudy}

As illustrated in Figure~\ref{fig:overview}, the \approach{} query phase consists of two main steps: candidate generation and ranking. In this experiment, we evaluate \approach{}'s performance on the candidate generation task based on the ground truth described in Section~\ref{subsec:ground_truth}, limited to those cases where a query entity has more than $10$ clicked target events.

As described in Section~\ref{sec:candidate_generation}, given a query entity $e$, the candidate generation step retrieves a set of candidate events regarding their embedding similarity towards $e$. To demonstrate the effectiveness of \approach{}'s language-specific embeddings for candidate generation, we compare the performance of different embedding methods. Here, we use the embedding techniques introduced in Section~\ref{sec:embeddings}.

For each query entity in the ground truth and each embedding technique, we retrieve the $200$ most similar events as candidate events. Then, we compute the candidate recall per embedding technique, i.e., the fraction of events in the ground truth contained in the candidate events. The results are shown in Table~\ref{tab:evaluation_embedding}. The DeepWalk and Node2Vec embeddings clearly outperform the other two embeddings, with the non-language-specific TransE embedding performing worst. This result demonstrates the benefit of creating language-specific random-walk-based embeddings for the language-specific event recommendation.

\begin{table}[]
\caption{Candidate Recall achieved by the \approach{} using different embedding methods.
}
\begin{tabular}{@{}lrrrr@{}}
\toprule
& \multicolumn{3}{c}{\textbf{Candidate Recall}} \\
\multicolumn{1}{c}{\textbf{Model}}   & \multicolumn{1}{c}{\textbf{German}} & \textbf{French} & \textbf{Russian} & \textbf{Avg.} \\ 
\midrule
\textbf{Deepwalk}                &           \textbf{0.408}                           &         \textbf{0.312}        & \textbf{0.373}    & \textbf{0.364}      \\ 
\textbf{Node2Vec}                &           0.348                          &         0.276        & 0.371     & 0.332       \\ 
\textbf{TransE}                &        0.009      & 0.007   & 0.008      & 0.008      \\
\textbf{Wikipedia2Vec}      &   0.017&    0.017          & 0.018     & 0.017       \\ 
\bottomrule 
\end{tabular}
\label{tab:evaluation_embedding}
\end{table}

\subsection{Recommendation Evaluation}
\label{subsec:ranking_study}

In this experiment, we evaluate \approach{}'s performance on the recommendation task based on the ground truth described in Section~\ref{subsec:ground_truth}.
Given a query entity and a set of candidate events, the goal in this task is to rank the candidate events according to their relevance to the query entity for the audience speaking the language of interest, i.e., the language-specific relevance.

In this experiment, \approach{} is trained via a $5$–fold cross-validation on each language separately. The folds are created based on the set of query entities: in each run, we use $80\%$ of the query entities and their events in the ground truth for training the LTR model, the remainder for testing. The results are averaged over the $5$ runs.

Table~\ref{tab:evaluation_reranking} reports the nDCG@10 and MAP@10 scores of the recommendation evaluation for the four recommendation baselines and \approach{} in three languages. As we can observe, in all the three languages, \approach{} clearly outperforms the baselines. On average, across languages, with an nDCG@10 of $0.957$ and MAP@10 of $0.97$, \approach{} outperforms the baselines by more than $8$ and $17$ percentage points, respectively. 
The \approach{} performance is similar across languages.

\begin{table*}[ht]
\center
\caption{nDCG@10 and MAP@10 scores achieved by the \approach{} approach and the recommendation baselines in three languages in the ranking study.
}
\begin{tabular}{@{}lrrrrrrrr@{}}
\toprule
 & \multicolumn{4}{c}{\textbf{nDCG@10 Score}}&\multicolumn{4}{c}{\textbf{MAP@10 Score}} \\
 & \multicolumn{1}{c}{\textbf{German}} & \multicolumn{1}{c}{\textbf{French}} & \multicolumn{1}{c}{\textbf{Russian}}& \multicolumn{1}{c}{\textbf{Avg.}}  & \multicolumn{1}{c}{\textbf{German}} & \multicolumn{1}{c}{\textbf{French}} & \multicolumn{1}{c}{\textbf{Russian}}& \multicolumn{1}{c}{\textbf{Avg.}} \\ \midrule
\textbf{Milne-Witten} & 0.893 & 0.897 & 0.890 & 0.893 & 0.848 & 0.864 & 0.838 & 0.850\\
\textbf{Node2Vec} & 0.860 & 0.841 & 0.885 & 0.862& 0.729 & 0.679 & 0.803 & 0.737 \\
\textbf{DeepWalk} & 0.899 & 0.858 & 0.901 & 0.886& 0.731 & 0.850 & 0.848 & 0.810 \\
\textbf{SuperGAT} & 0.853 & 0.884 & 0.879 & 0.872& 0.824 & 0.806 & 0.780 & 0.803 \\
\textbf{LaSER} & \textbf{0.957} & \textbf{0.958} & \textbf{0.956} & \textbf{0.957}& \textbf{0.969} & \textbf{0.970} & \textbf{0.971} & \textbf{0.970} \\ 
\bottomrule
\end{tabular}
\label{tab:evaluation_reranking}
\end{table*}

\subsection{Feature Analysis}
\label{sec:feature-analysis}

\sa{To assess the effectiveness of specific feature groups in \approach{}, we perform a feature analysis. To this extent, we leave out one feature group at a time and measure the resulting performance regarding nDCG@10. The results are presented in Table \ref{tab:ablation_study}. As we can observe, each feature group contributes towards the \approach{} overall performance. The link-based features ($Features_{links}$) provide the highest contribution among the four feature groups, while the temporal features ($Features_{temporal}$) have the lowest impact.}
We observe similar effects of feature groups across all languages.

A relatively low contribution of the spatial and temporal features can be explained through the 
non-availability of these features for a large proportion of entities, as reported in Section \ref{sec:entities-eventkg}.
Furthermore, whereas language-specific embeddings provide a substantial contribution in the candidate generation step, as discussed in Section \ref{subsec:embeddingstudy}, they have only a limited impact on the follow-up ranking step.
An average embedding-based similarity in this step is $0.65$ with a relatively low standard deviation of $\sigma=0.18$. Thus, re-ranking candidates based on the embedding-based similarity is only possible to a limited extent. 
Overall, incorporating all the proposed feature groups leads to the best performance of the proposed approach.

\begin{figure*}[ht]
    \centering
    \includegraphics[width=0.6\textwidth]{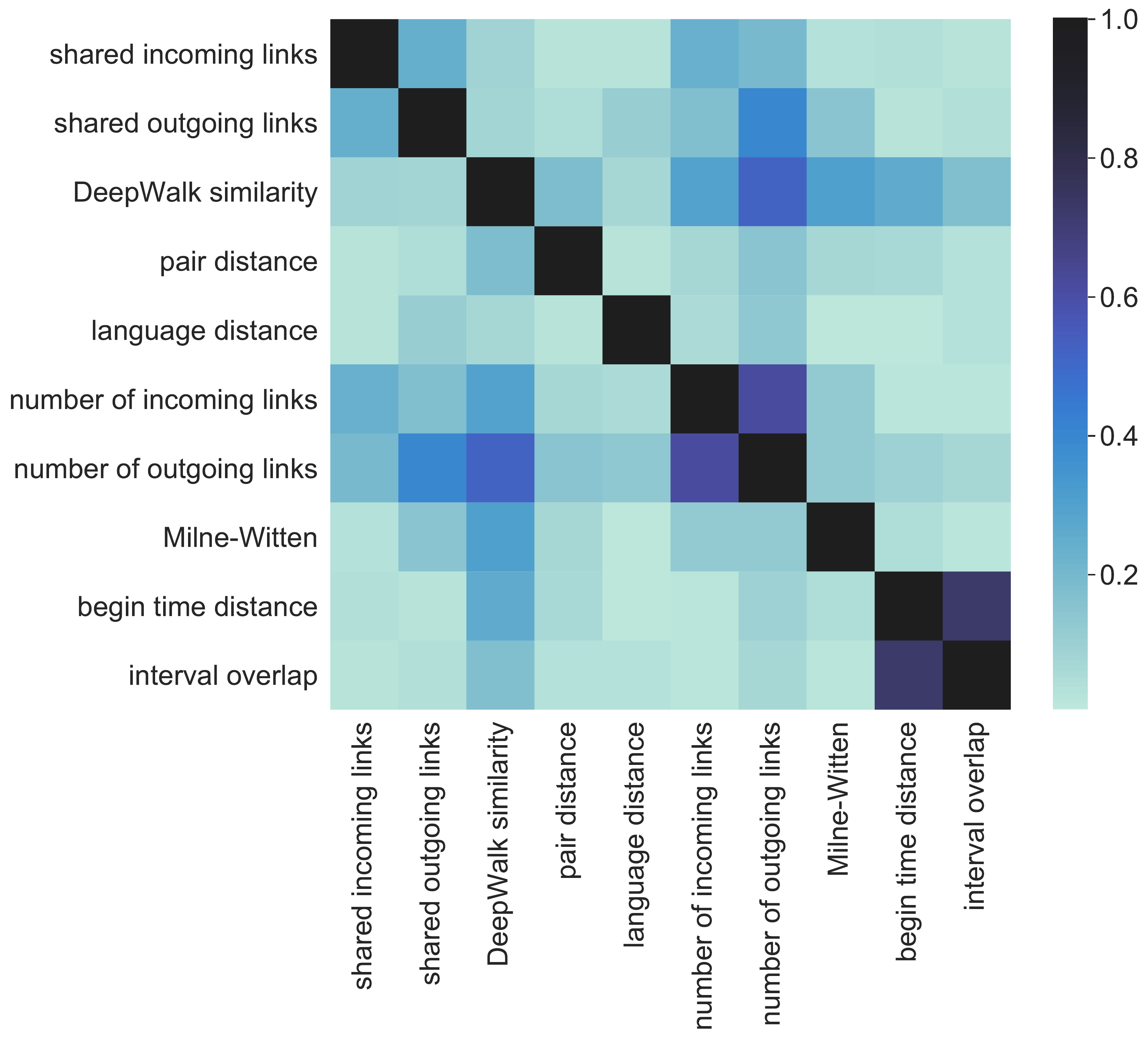}
    \caption{Feature correlation analysis using Pearson Correlation Coefficient (PCC). We report absolute correlation values.}\label{fig:feature-correlation}
\end{figure*}

Figure \ref{fig:feature-correlation} presents the correlation analysis of the features proposed in this article, computed using Pearson Correlation Coefficient (PCC). We report absolute correlation values. 
As we can observe, the highest correlation is obtained between the DeepWalk similarity and the number of links, as well as between the specific features in the time category. These correlations are expected.
Overall, we observe that the general feature categories provide complementary information with low correlation scores across categories.

\begin{table}[ht]
\center
\caption{Feature analysis: The results of \approach{} by leaving out feature groups. We report the nDCG@10 scores in three languages.}
\label{tab:ablation_study}
\begin{tabular}{llcc}
\toprule
\multicolumn{1}{c}{\textbf{Model}} & \multicolumn{1}{c}{\textbf{German}} & \textbf{French}      & \textbf{Russian}     \\
\midrule
\textbf{LaSER}  & 0.957 & 0.958 & 0.956 \\
- \textbf{$Features_{spatial}$}      &   0.956                                  &   0.952                   & 0.955               \\
- \textbf{$Features_{temporal}$}     &   0.956                                  &  0.957                    & 0.956    \\  
- \textbf{$Features_{links}$}        &   0.911                                  &   0.946                   & 0.909               \\
- \textbf{$Features_{embeddings}$}    & 0.950                                    &     0.957                 & 0.951               \\
\bottomrule
\end{tabular}
\end{table}

\section{User Study}
\label{sec:user_study}

The aim of the user study is to assess the recommendation quality from the user perspective. 
In this section, we describe the user study setup and discuss the results. 
Furthermore, we report the user agreement and provide insights into the user feedback.

\subsection{User Study Setup}
\label{subsec:user_study_setup}

Existing datasets such as the Wikipedia Clickstream usually only cover a fraction of events potentially relevant to a query entity. Consequently, evaluation is typically performed via pooling~\cite{ni2020layered}, i.e., by judging the relevance of the top recommendations generated by the methods under consideration \cite{zhang2016probabilistic}. In the user study, we followed the pooling approach to obtain the relevance judgments.

To conduct the user study, we selected $10$ popular query entities of various types, namely events, places, persons, art and religion. 
For each query entity, we generated event recommendations for German, French and Russian languages using the following methods: 
\ed{
\begin{itemize}
    \item The link-based Milne-Witten recommendation baseline, the best performing baseline according to Table \ref{tab:evaluation_reranking}. 
    \item The embedding-based DeepWalk recommendation baseline, the second best performing baseline according to Table \ref{tab:evaluation_reranking}.
    \item The proposed \approach{} approach. 
\end{itemize}
}

From each ranking, we select the top-$5$ highest ranked events and generate a set of $415$ (query entity, event, language) triples\footnote{As different approaches can recommend the same event, the total number of triples is less than $10$ (query entities) $\cdot 5$ (events per recommendation) $\cdot 3$ (languages) $\cdot 3$ (methods) $= 450$.}. 
To alleviate possible ranking-based bias in the judgments, we randomized the order of recommendations obtained by different methods for each query entity when presenting the recommendations to the study participants. 
Each triple was annotated by at least two study participants.

To gain more detailed insights into how the study participants perceive event relevance in a specific context, we break down the judgment into three \textit{relevance criteria}: (i) relevance to the topic (i.e., query entity), (ii) relevance to the audience of a language community and (iii) relevance to the general audience.

Following the TREC annotation guidelines~\cite{voorhees2002overview}, we asked the study participants to assume they want to write a report about the given topic and provide their relevance judgments regarding that setting. More specifically, we provided the participants with the following instructions: 

\begin{itemize}
\item Assume you want to write a report on the given topic (i.e. query entity)\footnote{For easier comprehension for the user study participants, we use the term ``topic'' in analogy to ``query entity'' in the user study interface.}. \begin{itemize}
\item Relevance to the topic: To what extent do you find the recommended event relevant to the topic and worth mentioning in your report?
\end{itemize}
\item Assume you want to write a report on the recommended event (independent of the given topic).\begin{itemize}
\item Relevance to the audience of a language community: To what extent do you find the event relevant to the audience that speaks the language?
\item Relevance to the general audience: To what extent do you find the event relevant to the general audience?\end{itemize}
\end{itemize}

For each recommended event and relevance criterion, the participants are asked to indicate whether the event is strongly relevant ($3$), partially relevant ($2$) or irrelevant ($1$).
Alternatively, the participants can select the ``I don't know'' option. 
To determine whether an event is relevant to a query entity in a specific language, we average over the user judgment scores.   
Events that exceed an average of $1.5$ are considered relevant when measuring precision of recommendations. Events judged as ``I don’t know'' by any of the study participants are excluded from the evaluation.

\begin{figure*}[t]
\includegraphics[width=\textwidth]{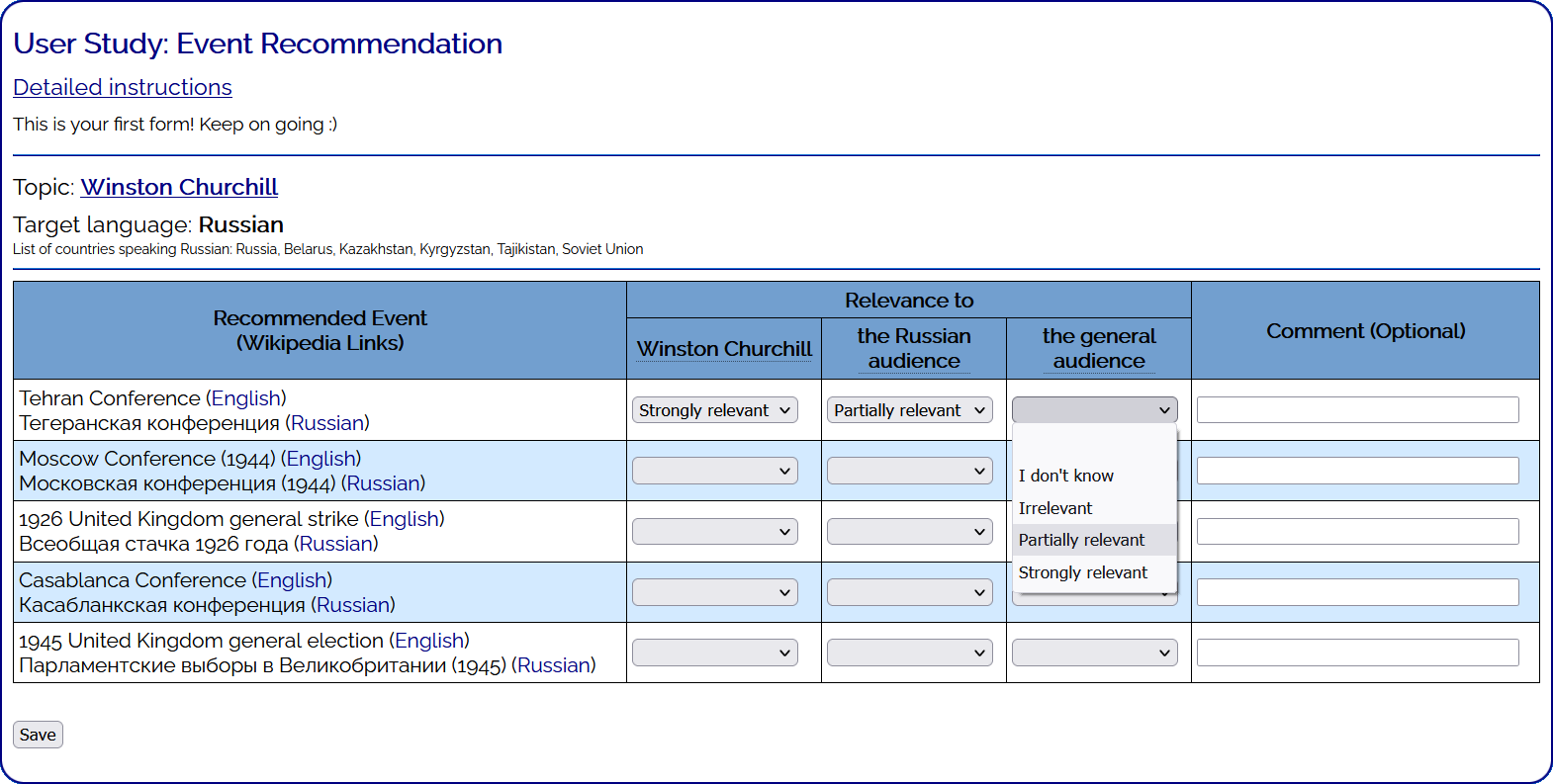}
\caption{Screenshot of the user study. Given a topic, i.e., a query entity (here: \textit{Winston Churchill}) and a target language (here: Russian), the participants are asked to judge the relevance of the recommended events. The participants assess each recommended event according to three relevance criteria: relevance to the topic, relevance to the audience of a specific language community, and relevance to the general audience. The relevance scores are as follows: strongly relevant, partially relevant, irrelevant, and unknown. An interface simultaneously presents five events (e.g., \textit{Tehran Conference}, \textit{Moscow Conference (1944)}, and others).}
\label{fig:study_screen}
\end{figure*}

A screenshot of the user study interface is presented in Figure~\ref{fig:study_screen}. The interface also provides links to the Wikipedia articles of the query entity and the recommended events, such that the participants can obtain additional information if required.

To collect feedback from the user study participants, they were given an option to leave a comment for each of their judgments. In addition, we conducted post-study interviews with selected participants.

As stated in Definition~\ref{def:laser}, we are interested in the relevance of an event to the query entity for the audience speaking the language of interest. 
Therefore, we derive language-specific relevance scores by requiring that both the relevance to the topic and the relevance to the language audience criteria are fulfilled.

\begin{itemize}
    \item Language-specific relevance: An event is language-specifically relevant when it is relevant to the topic (i.e., query entity) and relevant to the audience of the language community.
\end{itemize}

$17$ post-graduate researchers in Computer Science and Digital Humanities participated in the study. In total, participants annotated $935$ triples. Each participant annotated at least $9$ triples and $55$ triples on average.
The ratings are available online\footnote{\url{https://zenodo.org/record/5735580}}.

\subsection{User Study Results}
\label{sec:user-study-results}

Table~\ref{tab:userstudy_overall_precision} reports the MAP@5 scores for all languages and relevance criteria. As we can observe, \approach{} outperforms the recommendation baselines in the majority of the relevance criteria across the three languages. Most importantly, \approach{} obtains the highest MAP@5 on relevance to the language audience in all languages (French: $0.95$, German: $0.91$ and Russian: $0.84$) and outperforms the baselines by $10$ (German), $17$ (French) and $25$ percentage points (Russian). For German and French, \approach{} also achieves the highest relevance towards the general audience. Regarding the topic relevance, all approaches achieve MAP@5 scores larger than $0.9$ and \approach{} is slightly outperformed by the two baselines. \ed{This result is expected, as \approach{} aims at language-specific rather than generic topical relevance.}

To estimate \approach{}'s ability to provide recommendations which cover relevance to the topic and the language community, we are particularly interested in the language-specific relevance. Here, \approach{} outperforms both baselines in all languages. While the language-specific relevance for German is $0.81$ which is similar to Milne-Witten, the language-specific relevance for French and Russian is $0.90$ and $0.84$ and clearly exceeds those of the baselines by up to $33$ percentage points.

\begin{table*}[ht]
\center
 \caption{User study results: MAP@5 of \approach{} and two recommendation baselines in three languages regarding three relevance criteria judged in the user study and the overall language-specific relevance.}
    \label{tab:userstudy_overall_precision}
    \begin{tabular}{lcccc}
        \toprule
         
          & \multicolumn{3}{c}{\textbf{Relevance (MAP@5)}} & \multicolumn{1}{c}{\multirow{2}{*}{\textbf{\begin{tabular}[c]{@{}c@{}}Language-Specific\\ Relevance\end{tabular}}}} \\
 & \multicolumn{1}{c}{\textbf{Topic}} & \multicolumn{1}{c}{\textbf{General Audience}} & \multicolumn{1}{c}{\textbf{Language Audience}} & \multicolumn{1}{c}{} \\ \midrule
         
        \multicolumn{5}{c}{\textbf{German}} \\ 
        Milne-Witten & \textbf{1.00} & 0.87 & 0.81 & \textbf{0.81} \\ 
        DeepWalk & 0.98 & 0.77 & 0.70 & 0.70 \\ 
        \approach{} & 0.93 & \textbf{0.89} & \textbf{0.91} & \textbf{0.81} \\ \midrule
        \multicolumn{5}{c}{\textbf{French}} \\ 
        Milne-Witten & \textbf{1.00} & 0.88 & 0.68 & 0.68 \\ 
        DeepWalk & \textbf{1.00} & 0.88 & 0.78 & 0.77 \\ 
        \approach{} & 0.95 & \textbf{0.94} & \textbf{0.95} & \textbf{0.90} \\ \midrule
        \multicolumn{5}{c}{\textbf{Russian}} \\ 
        Milne-Witten & \textbf{1.00} & 0.90 & 0.59 & 0.59 \\ 
        DeepWalk & \textbf{1.00} & \textbf{0.95} & 0.51 & 0.51 \\ 
        \approach{} & 0.98 & 0.90 & \textbf{0.84} & \textbf{0.84} \\ \bottomrule
\end{tabular}%

\end{table*}

Table~\ref{tab:userstudy_detailed_precision} presents \approach{}'s performance for each of the query entities annotated in the user study separately. The highest AP@5 scores considering the relevance to the language community are achieved for query entities of type \textit{Event} including sports and cultural events. Even in cases where the query entity has general engagement and impact, such as \textit{Olympic Games} and \textit{World War II}, our proposed approach can recommend specific events to each language community. 

\begin{table*}[]
\center
\caption{User Study: Detailed analysis of \approach{} for all query entities annotated in the user study. For each query entity and a language, we report the average precision (AP@5) for the three relevance criteria (Topic: Relevance to the Query Entity, Lang.: Relevance to the audience of a language community, General: relevance to the general audience).}
\label{tab:userstudy_detailed_precision}
\begin{tabular}{llrrrrrrrrr}
\toprule
 & \textbf{} & \multicolumn{3}{c}{\textbf{AP@5 (German)}} & \multicolumn{3}{c}{\textbf{AP@5 (French)}} & \multicolumn{3}{c}{\textbf{AP@5 (Russian)}} \\ 
 \textbf{Query Entity} & \textbf{Type} & \textbf{Topic} & \textbf{Lang.} & \textbf{General} & \textbf{Topic} & \textbf{Lang.} & \textbf{General} & \textbf{Topic} & \textbf{Lang.} & \textbf{General} \\ \midrule
Christianity & Religion & 1 & 1 & 1 & 1 & 1 & 1 & 0.8 & 1 & 0.91 \\ 
Film festival & Cultural event & 1 & 1 & 1 & 1 & 1 & 1 & 1 & 1 & 1 \\
Germany & Place & 1 & 1 & 1 & 1 & 1 & 1 & 0.8 & 1 & 0.54 \\ 
Olympic Games & Sport event & 1 & 1 & 1 & 1 & 1 & 1 & 1 & 1 & 1 \\ 
Painting & Art & 0.8 & 0.54 & 0.91 & 1 & 1 & 1 & 0.2 & 0.8 & 0.8 \\ 
Social movement & Group action & 1 & 0.91 & 1 & 1 & 1 & 0.84 & 0.8 & 1 & 0.84 \\ 
UK & Place & 0.6 & 0.55 & 0.61 & 1 & 0.50 & 0.54 & 0.8 & 1 & 0.96 \\ 
Winston Churchill & Person & 0.29 & 0.96 & 0.84 & 1 & 1 & 1 & 0.8 & 0.8 & 0.96\\
World War I & Event & 0.87 & 0.96 & 0.25 & 0.45 & 1 & 1 & 1 & 1 & 1 \\ 
World War II & Event & 1 & 1 & 1 & 1 & 1 & 1 & 1 & 1 & 1 \\ \bottomrule
\end{tabular}%
\end{table*}

\subsection{User Agreement}

To estimate the difficulty of providing relevance judgments for language-specific event recommendations, we measure the agreement between the user study participants using Fleiss's kappa statistic. This statistic measures agreement among any constant number of raters \cite{fleiss1971measuring}, where the values less than $0$ indicate a poor agreement and $1$ a perfect agreement.

To assess the user agreement, we conducted a second phase of the user study following the same setup, but under the following conditions:
\begin{itemize}
    \item We considered \textit{Film festival} as the query entity. 
    \item We collected judgments from $5$ users for each (query entity, event, language) triple.
    \item For each language, we added $5$ negative examples randomly selected from the set of all events.
\end{itemize}

In total, we collected $270$ annotations for $54$ (query entity, event, language) triples in the user agreement study.

We compute the agreement among the study participants for evaluating the events in three languages, three different relevance criteria and three classes (partially relevant, strongly relevant and irrelevant). The resulting Fleiss’s kappa ($\kappa$) values are shown in Table~\ref{tab:evaluation_user_agreement}.
According to these results, users achieved a substantial agreement when judging whether a recommended event is relevant to the query entity ($\kappa \geq 0.62$). We observe a moderate agreement for relevance to the language audience (average $\kappa = 0.44$).
Interestingly, the users disagreed the most when judging the relevance of an event to the general audience (average $\kappa = -0.01$). To check if users at least agreed on the irrelevance of events, we also computed Fleiss’s kappa considering only two classes (``partially or strongly relevant'' and ``irrelevant'') for the relevance to the general audience, which slightly improved the measured agreement (average $\kappa = 0.16$).

The user agreement results confirm that the consideration of language-specificity is important, as this dimension can be captured more easily than the general relevance by the users.

\begin{table}[]
\setlength\tabcolsep{1.8pt}
\center
\small
\caption{Inter-rater agreement assessment using Fleiss' kappa.}

\label{tab:evaluation_user_agreement}
\begin{tabular}{lrrrr}
\toprule
                   \multicolumn{1}{l}{\textbf{\begin{tabular}[l]{@{}l@{}}Agreement on\\ Relevance to the\end{tabular}}}                         & \multicolumn{1}{c}{\textbf{German}}    & \multicolumn{1}{c}{\textbf{French}}    & \multicolumn{1}{c}{\textbf{Russian}} & \multicolumn{1}{c}{\textbf{Avg.}}   \\ \midrule
\textbf{Topic}                          & 0.62 & 0.79 & 0.89 & 0.78 \\
\textbf{Language audience}                       & 0.39 & 0.48 & 0.44 & 0.44\\
\textbf{General audience}                & 0.01 & 0.02 & -0.05 & -0.01 \\
\textbf{General audience (binary)} & 0.15 & 0.10 & 0.23 & 0.16 \\ \bottomrule
\end{tabular}
\end{table}

There were cases where the same event was annotated with all grades of relevance (strongly, partially and irrelevant), specifically when judging relevance to the general audience. Examples include \textit{46th Venice International Film Festival} and \textit{All-Union Film Festival}. While some users see specific film festivals as generally relevant, others disagree. These examples demonstrate the subjectivity when providing relevance judgments. For example, a film fan might be more convinced to rate specific film festivals as generally relevant than others.


\subsection{Participant Feedback}

To gain further insights into the relevance judgments provided by the study participants, we now look at their feedback, which was collected from the comments they could provide during the study (see Figure~\ref{fig:study_screen}) and in post-study interviews with selected participants. From this feedback, we identify the following challenges the study participants were facing:

\begin{itemize}
\item Lack of information: In some cases, study participants were unable to retrieve enough information about a recommended event to make a confident judgment. Examples include the triple (\textit{World War I}, \textit{Colmar Pocket}, French) (user comment: ``I have heard nothing about that'') and (\textit{Winston Churchill}, \textit{Litvinov Protocol}, Russian) (user comment: ``I tried to find this word in the English Wikipedia, but I did not find it there.''). 
This example also illustrates the information asymmetry across the Wikipedia language editions.

\item Difficulty in judging the relevance to the general audience: In a post-study interview, one study participant described that it was difficult to decide on the relevance to the general audience. Regarding relevance to the language audience, the user identified more intuitive criteria, such as the location of the event or its participants. In contrast, the user struggled with finding particular criteria to measure the relevance to the general audience.

This observation confirms our interpretation of the user agreement analysis, where we observed only a slight agreement regarding the relevance for the general audience.

\item Wrong classification: In a few cases, the recommendation was not an event. An example is the triple (\textit{Christianity}, \textit{Constantinople}, French) (user comment: ``Not an event''). This error type can be explained by wrong event type assignments in the underlying knowledge graph. In our evaluation, those cases are excluded if one of the users selected the ``I don't know'' option.
\end{itemize}

From the participant feedback, we learn that the distinction into three relevance scores helps guide the user study participants through their annotations, specifically in the cases where users do not feel confident with their judgment due to missing information.

\section{Anecdotal Results and Application Scenarios}
\label{sec:anecdotal_results}

In our final evaluation step, we analyze selected event recommendations of \approach{} for two query entities annotated during the user study. In this section, we present anecdotal results and application scenarios to highlight the strengths of the proposed \approach{} approach.

\subsection{Anecdotal Example 1: Film Festival}
As our first example query entity, we select \textit{Film Festival} as a rather generic topic. The top-5 events recommended by \approach{} for German, French and Russian are shown in Table~\ref{tab:film_festival_results}. The recommended events clearly show a language-specific focus, i.e., important film festivals that happened in cities where the respective languages are spoken: for example, \approach{} recommends \textit{International Short Film Festival Oberhausen} and \textit{Filmfest München} for the German audience, several César Awards and  \textit{Brest European Short Film Festival} for the French audience and \textit{Moscow International Film Festival} as well as the \textit{\foreignlanguage{russian}{Окно в Европу}}, a Russian film festival happening in the Russian city \textit{Wyborg}, for the Russian audience.

\subsection{Anecdotal Example 2: \textit{World War II}}

As another query entity, we selected \textit{World War II} as a concrete event for which we expect to retrieve a lot of regionally significant sub-events. The generated recommendations are shown in Table~\ref{tab:world_war_II_results}. As expected, these recommendations contain operations, battles, and conferences that happened in the immediate context of \textit{World War II}. Again, the recommended events show a language-specific focus: For German, we get \textit{Aktion Silberstreif}, \textit{Battle of Loos} in the French city \textit{Loos} for French and \textit{\foreignlanguage{russian}{Бои в Яновских лесах}} (Battles in the Janowska forests) for Russian. Other recommended events such as \textit{Vietnamese famine of 1945}, \textit{Spa Conference of 1920} and \textit{Death of Adolf Hitler} show a less language-specific focus. However, they represent important happenings during \textit{World War II}.

In general, the two presented anecdotal examples further illustrate that \approach{} can recommend events that differ across languages, reflecting language-specific relevance.

\begin{table}[ht]
\caption{Anecdotal example 1: Events recommended for the query entity \textit{Film Festival} in three languages.}
\label{tab:film_festival_results}
\begin{tabular}{lll}
\\ \toprule
\multicolumn{3}{c}{\textbf{Film Festival}}                                                                    \\ \midrule
\multicolumn{1}{c}{\multirow{5}{*}{German}} & 1 & 46th Venice International Film Festival            \\
\multicolumn{1}{c}{}                        & 2 & International Short Film Festival Oberhausen       \\
\multicolumn{1}{c}{}                        
                                            & 3 & \begin{tabular}[c]{@{}l@{}}KALIBER35 Munich International \\ Short Film Festival\end{tabular} \\

\multicolumn{1}{c}{}                        & 4 & Filmfest Hamburg                                   \\
\multicolumn{1}{c}{}                        & 5 & Filmfest München                                   \\ \hline
\multirow{5}{*}{French}                     & 1 & 34th César Awards                                  \\
                                            & 2 & 6th César Awards                                   \\
                                            & 3 & 21st Lumières Awards                               \\
                                            & 4 & 17th César Awards                                  \\
                                            & 5 & Brest European Short Film Festival                 \\ \hline
\multirow{6}{*}{Russian}                    & 1 & All-Union Film Festival                            \\
                                            & 2 & \begin{tabular}[c]{@{}l@{}}\foreignlanguage{russian}{Окно в Европу (кинофестиваль)} \\ \textit{Window to Europe (film festival)}\end{tabular} \\
                                            & 3 & {Moscow International Film Festival}                 \\
                                            & 4 & {Short Film}                                         \\
                                            & 5 & {Kinotavr}                                           \\ \bottomrule
\end{tabular}
\end{table}

\begin{table}[tb]
\caption{Anecdotal example 2: Events recommended for the query entity \textit{World War II} in three languages.}
\label{tab:world_war_II_results}
\begin{tabular}{lll}
\toprule 
\multicolumn{3}{c}{\textbf{World War II}}                                                              \\ \midrule 
\multicolumn{1}{c}{\multirow{5}{*}{German}} & 1 & Aktion Silberstreif                         \\
\multicolumn{1}{c}{}                        & 2 & Operation Jedburgh                          \\
\multicolumn{1}{c}{}                        & 3 & Vietnamese famine of 1945                   \\
\multicolumn{1}{c}{}                        & 4 & Einsatzgruppen trial                        \\
\multicolumn{1}{c}{}                        & 5 & Operation Felix                             \\ \hline
\multirow{5}{*}{French}                     & 1 & Spa Conference of 1920                      \\
                                            & 2 & French war planning 1920–1940               \\
                                            & 3 & Locarno Treaties                            \\
                                            & 4 & Battle of Loos                              \\
                                            & 5 & Treaty of Neuilly-sur-Seine                 \\ \hline
\multirow{7}{*}{Russian}                   
                                            & 1 & \begin{tabular}[c]{@{}l@{}}\foreignlanguage{russian}{Бои в Яновских лесах} \\ \textit{Battles in the Janowska forests}\end{tabular} \\
                                            & 2 & \begin{tabular}[c]{@{}l@{}}\foreignlanguage{russian}{События в Польше в сентябре 1939 года} \\ \textit{Events in Poland in September 1939}\end{tabular} \\
                                            & 3 & \begin{tabular}[c]{@{}l@{}}\foreignlanguage{russian}{Гибель авианосца «Глориес»} \\ \textit{Sinking of the aircraft carrier Glorious}\end{tabular} \\
                                            & 4 & Defence of the Polish Post Office in Danzig \\
                                            & 5 & Death of Adolf Hitler                       \\ \bottomrule
\end{tabular}

\end{table}

\subsection{Application Scenarios}

Motivated by the anecdotal examples, we envision the following application scenarios based on language-specific event recommendations:

\begin{itemize}
    \item \textbf{Within-language exploration}: Exploration of events related to a topic in a specific language can strengthen the exploration focus. For example, a historian researching the course of \textit{World War II} in France might be specifically interested in relevant happenings related to France, including the French war planning and \textit{Battle of Loos} shown in Table~\ref{tab:world_war_II_results}. With \approach{}, such historian could easily collect such events and use them as a basis for further research.
      \item \textbf{Cross-language exploration}: Language-specific event recommendation could also help to explore topics from a variety of viewpoints and thus widen horizons and minimize potential cultural biases which are prevalent on the Web \cite{nangia2020crows}. A user can specify multiple languages and explore the respective events related to a topic. In our \textit{Film Festival} example in Table~\ref{tab:film_festival_results}, this procedure would result in a collection of important film festivals in different parts of the world. That way, one can even explore events such as \textit{\foreignlanguage{russian}{Окно в Европу} (Window to Europe)}, which are only described in a specific language but might be of interest for the cross-language exploration.
    \item \textbf{User language background}: The language background is often part of a user's profile. Thus, recommending events specifically relevant to the user language can potentially satisfy the user-specific information needs and increase user satisfaction in web navigation and exploratory search.
\end{itemize}


\section{Related Work}
\label{sec:related_work}

The \approach{} approach presented in this article aims at recommending events by taking their language-specific relevance into account. 
This section describes \sg{related research areas including entity and event recommendation, learning to rank,} graph embeddings, and cross-lingual research.

\subsection{Recommendation}

While \sg{the tasks of user-item and entity recommendation have} been extensively studied in the literature, event recommendation has until now been mainly limited to social media events.

\subsubsection{User-Item Recommendation}

A typical recommendation task is that of user-item recommendation, where items (e.g., movies or points of interest) are recommended to an individual user \cite{yu2016network}, typically based on a network of users, items, and their interactions. The recommendation can be based on the user's past preferences, taking into account the preferences of similar users (collaborative filtering) or the similarity to other items (content-based filtering). Knowledge graphs have been used to serve as background knowledge for item-user recommendations where they provide additional information regarding the connections between items \cite{guo2020survey, wang2019kgat, wang2020kerl, wang2021learning}. 

The task of language-specific event recommendation introduced in this article has different prerequisites than the user-item recommendation and is thus not comparable. Event recommendations are provided given a query entity, a query language, and a language-specific knowledge graph. However, there is no user-item network and consequently no possibility to incorporate the preferences of individual users.

\subsubsection{Entity Recommendation}
Entity recommendation is the task of recommending a ranked list of entities to the user query. Blanco et al.~\cite{blanco2013entity} presented Spark, an entity recommendation system that, by using and combining several signals from a variety of data sources, provides a ranking of the entities related to the user query. 
Ni et al. \cite{ni2020layered} proposed a framework for recommending related Wikipedia entities using an architecture of multiple layered graphs, candidate generation via Doc2Vec embeddings, and ranking with an LTR model.
In contrast to \approach{}, this model is trained on proprietary search log data and is therefore difficult to reproduce.
Other approaches focus on specific recommendation aspects: 
Zhang et al.~\cite{zhang2016probabilistic} proposed a time-aware entity recommendation (TER), which allows users to restrict their interests in entities to a customized time range. Tran et al.~\cite{tran2017beyond} extended TER by incorporating topic and time and proposed contextual relatedness among entities using embedding techniques. The user interest and preference have also been studied by Bi et al.~\cite{bi2015learning} who proposed ``probabilistic Three-way Entity Model'' (TEM) that provides personalized recommendations of related entities using user interactions from personal click logs. Huang et al.~\cite{huang2018learning} studied serendipity to engage the interest of users while recommending entities.

Existing entity recommendation methods are focused on recommendation regardless of language preferences. Unlike these methods, our proposed \approach{} approach takes languages into account and recommends relevant language-specific events. 

\subsubsection{Event Recommendation}

Events take an important role in a range of real-world applications, including news search~\cite{rudnik2019searching}, news linking~\cite{setty2018event2vec} and event-centric user interfaces~\cite{gottschalk2018eventkg}. However, event recommendation has not been extensively studied and is primarily focused on social media events. Existing event recommendation approaches \cite{gao2016collaborative} focus on event-based social networks (ESRN) such as Meetup, where the goal is to recommend social events such as parties, concerts, and conferences to the users. 
Unlike the approaches mentioned above, we focus on events of societal importance, such as the Coronavirus pandemic and the Second World War. With the proposed \approach{} approach, we leverage structured information from knowledge graphs and consider information needs and the specific context of language communities. 

\subsection{Learning to Rank}

The ranking is an essential step of many recommendation algorithms, typically following the candidate generation step. Given a set of objects, a ranking model calculates the score of each object and sorts them accordingly. The scores may represent the degrees of relevance, preference, or importance, depending on applications \cite{liu2009learning}.

LambdaMART is a state-of-the-art LTR model that uses a boosted tree model, which is, according to \cite{guo2020deep}, still ''hard to beat for most neural ranking models based on raw texts``. 
LambdaMART demonstrated superior performance when click-based data were used as features \cite{wu2018turning} and has been applied in many application domains, including recommendations \cite{palumbo2017entity2rec}, e-commerce click and search \cite{ guo2020debiasing}. Our evaluation of \approach{} confirms LambdaMART's superiority for the task of language-specific event recommendation. 

\subsection{Embedding Methods}

Graph embedding techniques have been recently adopted for recommendation tasks~\cite{alghamdi2021learning}. Graph embeddings aim to represent graph nodes by low-dimensional vectors, which preserve the graph structure. They are created using random-walk-based, deep-learning-based, and factorization-based methods. Random-walk-based methods \cite{perozzi2014deepwalk, grover2016node2vec} capture the node neighborhoods by creating random walks over the graph nodes fed into language models in analogy to sentences. Such methods are beneficial for large graphs when the graph is too large to cover in its entirety \cite{goyal2018graph}. Well-known random-walk-based methods include DeepWalk~\cite{perozzi2014deepwalk} and Node2Vec~\cite{grover2016node2vec} which learn continuous feature representations for nodes based on biased random walks that provide a trade-off between breadth-first and depth-first graph search. Graph embedding methods based on deep learning \cite{cao2016deep} typically learn auto-encoders to compress information about the local node neighborhood~\cite{hamilton2017representation}. 
Factorization-based algorithms \cite{ou2016asymmetric} represent the connections between nodes in the matrix form and factorize this matrix to obtain the embedding \cite{goyal2018graph}.
Due to the large size of the \lkg{} we employ random-walk-based graph embedding methods, which provide an effective solution for large graphs.

Knowledge graph embeddings specifically target the embedding of entities and relations in a knowledge graph and are used for knowledge graph completion, relation extraction, and other tasks. Translational distance models such as TransE \cite{bordes2013translating} and its extensions exploit distance-based scoring functions. They measure the plausibility of a fact as the distance between the two entities, usually after a translation carried out by the relation~\cite{wang2017knowledge}. Other knowledge graph embedding methods employ additional information such as entity types \cite{xie2016representation}, relation paths \cite{toutanova2016compositional}, textual information \cite{xie2016representation} and hybrid information (e.g., Wikipedia2Vec \cite{yamada2020wikipedia2vec}) in the embedding process. In Section~\ref{subsec:embeddingstudy}, we discuss the impact of different knowledge graph embedding methods on \approach{} and the benefits of using language-specific embeddings for the candidate generation.

 \subsection{Cross-lingual Research}

The emerging need to analyze multilingual information on the web has been targeted in a variety of studies, e.g., ~\cite{roy2021information}. Wikipedia is an essential source for multilingual studies regarding the content, number of users, and language coverage. In this regard, several studies focused on investigating and exploring cross-lingual differences in Wikipedia \cite{gottschalk2017multiwiki}. By analyzing bias and linguistic points of view regarding a controversial event, R. Rogers~\cite{rogers2013digital} illustrates that Wikipedia articles varied in their titles and the content across the different language editions. Other works studied multilingualism in terms of user editing behavior~\cite{hale2014multilinguals} and reflection of cross-cultural similarities in the process of collective archiving knowledge on Wikipedia~\cite{samoilenko2016linguistic}.

The \approach{} approach presented in this article provides an intuitive way to explore language-specific events that can be beneficial for language-specific and cross-lingual studies.

\section{Conclusion}
\label{sec:conclusion}

In this article, we defined the novel task of language-specific event recommendation. We presented \approach{}, a novel approach to tackle this task. \approach{} recommends a list of events relevant to the query entity in a language-specific context.
After the creation of language-specific entity embeddings, we train a learning to rank model that generalizes from \cg{} using spatial, temporal, link-based and latent embedding-based features. 
\sa{We experimentally demonstrate the benefit of creating language-specific embeddings for the task of language-specific event recommendation.}
Furthermore, our experiments on a real-world dataset demonstrate the effectiveness of \approach{} in ranking events compared to link-based, embedding-based and graph attention network-based recommendation baselines, outperforming them by more than $8$ (nDCG@10)  and $17$ (MAP@10) percentage points, respectively.
Moreover, we identified and analyzed different relevance criteria in a user study and demonstrated that \approach{} effectively recommends events in a language-specific context. 
Regarding the language-specific relevance, \approach{} outperforms the best performing baselines by up to $33$ percentage points in MAP@5. 
Our evaluation demonstrates that language-specific context is an essential event recommendation criterion, together with topical and global event relevance.

\subsubsection*{Acknowledgments} 
This work was partially funded by H2020-MSCA-ITN-2018-812997 under ``Cleopatra'' and
by DFG, German Research Foundation under ``WorldKG'' (424985896).






\balance

\bibliographystyle{elsarticle-num}           
\bibliography{bibliography}        

%

\end{document}